\documentclass[fleqn,usenatbib]{mnras}

\usepackage{CJKutf8}
\usepackage[T1]{fontenc}
\usepackage{hyperref}
\usepackage{xcolor}

\DeclareRobustCommand{\VAN}[3]{#2}
\let\VANthebibliography\thebibliography
\def\thebibliography{\DeclareRobustCommand{\VAN}[3]{##3}\VANthebibliography}

\usepackage{graphicx}	
\usepackage{amsmath}	
\usepackage{amssymb}	
\usepackage{multirow}
\usepackage{newtxtext,newtxmath}

\title[Complexity Scan of EMU-PS]{Identifying anomalous radio sources in the EMU Pilot Survey using a complexity-based approach}

\author[Gary Segal et al.]{Gary Segal,$^{1,2}$\thanks{E-mail: g.segal@uq.edu.au}
David Parkinson,$^{3}$\thanks{E-mail: davidparkinson@kasi.re.kr} 
Ray Norris,$^{2,4}$ 
Andrew M.~Hopkins,$^{5}$ 
Heinz Andernach,$^{6}$\thanks{Permanent address: Departamento de Astronom\'{i}a, Universidad de Guanajuato, Callej\'on de Jalisco s/n, Guanajuato, C.P. 36023, GTO, Mexico} 
\newauthor
Emma L. Alexander,$^{7}$ 
Ettore~Carretti,$^{8}$ 
B\"arbel S. Koribalski,$^{2,4}$ 
Letjatji S. Legodi,$^{9}$ 
\newauthor
Sarah Leslie,$^{10}$  
Yan Luo,$^{11}$
Jonathon C. S. Pierce,$^{12}$ 
Hongming Tang,$^{13}$ 
Eleni Vardoulaki,$^{14}$ 
\newauthor
Tessa Vernstrom$^{15}$ 
\\
$^{1}$School of Mathematics and Physics, University of Queensland, St Lucia, Brisbane, QLD 4072, Australia\\
$^{2}$CSIRO Space and Astronomy, PO Box 76, Epping, 1710, NSW, Australia\\
$^{3}$Korea Astronomy and Space Science Institute, Daejeon 34055, Korea\\
$^{4}$Western Sydney University, Locked Bag 1797, Penrith, NSW 2751, Australia\\
$^5$Australian Astronomical Optics, Macquarie University, 105 Delhi Rd, North Ryde, NSW 2113, Australia\\
$^{6}$Th\"uringer Landessternwarte, Sternwarte 5, D-07778 Tautenburg, Germany\\ 
$^{7}$Jodrell Bank Centre for Astrophysics, Department of Physics and Astronomy, University of Manchester, Manchester, UK\\
$^{8}$INAF, Istituto di Radioastronomia, Via Gobetti 101, 40129 Bologna, 
Italy\\
$^{9}$South African Radio Astronomy Observatory, 2 Fir Street, Black River Park, Observatory, Cape Town, 7925, South Africa\\
$^{10}$Leiden Observatory, Leiden University, PO Box 9513, NL-2300 RA Leiden, The Netherlands\\
$^{11}$School of Physics and Astronomy, Sun Yat-sen University, 2 Daxue Road, Zhuhai 519082, China\\
$^{12}$Centre for Astrophysics Research, University of Hertfordshire, College Lane, Hatfield AL10 9AB, UK\\
$^{13}$Department of Astronomy, Tsinghua University, Beijing 100084, China\\
$^{14}$Th\"{u}ringer Landessternwarte, Sternwarte 5, 07778 Tautenburg, Germany\\
$^{15}$ICRAR, The University of Western Australia, 35 Stirling Hwy, 6009 Crawley, Australia
}

\date{Accepted XXX. Received YYY; in original form ZZZ}

\pubyear{2022}

\begin{document}
\label{firstpage}
\pagerange{\pageref{firstpage}--\pageref{lastpage}}
\maketitle

\begin{abstract}

The Evolutionary Map of the Universe (EMU) large-area radio continuum survey will detect tens of millions of radio galaxies, giving an opportunity for the detection of previously unknown classes of objects. To maximise the scientific value and make new discoveries, the analysis of this data will need to go beyond simple visual inspection. We propose the coarse-grained complexity, a simple scalar quantity relating to the minimum description length of an image, that can be used to identify unusual structures. The complexity can be computed without reference to the broader sample or existing catalogue data, making the computation efficient on new surveys at very large scales (such as the full EMU survey). We apply our coarse-grained complexity measure to data from the EMU Pilot Survey to detect and confirm anomalous objects in this data set and produce an anomaly catalogue. Rather than work with existing catalogue data using a specific source detection algorithm,  we perform a blind scan of the area, computing the complexity using a sliding square aperture. The effectiveness of the complexity measure for identifying anomalous objects is evaluated using crowd-sourced labels generated via the Zooniverse.org platform. We find that the complexity scan identifies unusual sources, such as odd radio circles, 
by partitioning on complexity. We achieve partitions where 5\% of the data is estimated to be 86\% complete, and 0.5\% is estimated to be 94\% pure, with respect to anomalies and use this to produce an anomaly catalogue.

\end{abstract}




\section{Introduction}

The large-scale analysis of the extragalactic sky has, in the past, delighted astronomers with new and unusual objects. We have no doubt that it will continue to do so into the future, with new large-scale surveys such as the Legacy Survey of Space and Time \citep[LSST]{2019ApJ...873..111I}, Dark Energy Spectroscopic Instrument \citep[DESI]{2016arXiv161100036D}, Evolutionary Map of the Universe \citep[EMU]{2011PASA...28..215N}, LOFAR Two-metre Sky Survey \citep[LoTSS]{2022A&A...659A...1S,2017A&A...598A.104S}, MeerKAT International GHz Tiered Extragalactic Exploration Survey \citep[MIGHTEE]{jarvis16},  Spectro-Photometer for the History of the Universe, Epoch of Reionization, and Ices Explorer \citep[SPHEREx]{2014arXiv1412.4872D}, and the Square Kilometre Array \citep[SKA]{2009IEEEP..97.1482D}, either in operation or starting very soon. The work of the astrophysicist is to understand these objects, learn their nature, and identify if they fall inside some already understood class, or constitute an entirely new type of object. For objects that have features or attributes that are completely unexpected, so called `unknown unknowns', even detecting these in the first place may be a challenge \citep{norris2017}. In the present paper we define an anomaly, in broad terms, as an observation that is considered unexpected, based on consensus votes from astrophysicists.\footnote{Similar to the threshold suggested by the United States Supreme Court Justice Potter Stewart} While a human may easily notice something that is unexpected, training a machine to do so may be more difficult. The application of machine learning approaches to astrophysics, such as outlier detection, has already seen some developments with a number of different algorithms or approaches (e.g. \cite{mostert2021unveiling,2020MNRAS.499..524G, baron17}). The complexity based approach offers a more computationally efficient tool for anomaly detection based on the morphology of radio sources.

The effectiveness of other machine learning approaches, such as Convolutional Neural Networks, have been demonstrated for identifying and classifying observations in astronomical surveys based on their features \citep{Thorat,lukic2018radio,lukic2019morphological,karpenka,kimbailerjones,desspcc,Dieleman,Huertas-Company,Charnock,wu2019radio,lochner2016photometric}. Unsupervised learning approaches, such as self-organising maps, have also been applied to clustering and segmentation problems including PINK by \citet{polsterer2015automatic,polsterer2019pink} and applications by \citet{ baron17,galvin20,mostert2021unveiling,Nikhel}. \citet{lochner2021astronomaly} developed Astronomaly as a general anomaly detection framework based on an active learning approach that provides personalised recommendations. Astronomaly \citep{lochner2021astronomaly} was designed to work with a broad range of astronomical data from images to spectra. \citet{lochner2021astronomaly} used a Galaxy Zoo project to demonstrate the effectiveness of the approach, where Astronomaly was found to double the number of interesting objects found within the first 100 viewed within the datasets.

In our original paper (\citet{2019PASP..131j8007S}, hereafter S19), we introduced the idea of the coarse-grained complexity measure as a tool for identifying complex and anomalous objects. This quantity was based on the notions of \textit{effective complexity} defined by \citet{GellMann94,gell1996information} and \textit{apparent complexity} defined and implemented in \citet{Carroll}, as the information required to describe a system's regularities, or more specifically, the entropy approximated by an upper bound on the Kolmogorov complexity after applying a smoothing function.\footnote{Random intensity fluctuations (noise) cannot be  compressed and will increase the description length required to represent possible states beyond those generated by the source object. This noise does not describe the complexity of the object of interest and is reduced through smoothing as part of the measurement of the coarse-grained complexity.} In S19 we used data from the Australia Telescope Large Area Survey (ATLAS), to measure the coarse-grained complexity of radio continuum images using the \texttt{gzip} \citep{gzip} byte length, post smoothing, to estimate the upper bound of the complexity value.  We found it to be  useful (when combined with clustering methods to automate the process) for segmenting complex or unusual images from simple images without requiring large training data and without learning specific features from labelled data. The approach generalised well when applied to new data after being calibrated on a much smaller dataset, with implemented at worst-case linear time complexity.\footnote{The ``worst-case time complexity`` refers to an upper bound on the time to run an algorithm by counting elementary operations for all permissible inputs. The efficiency of the algorithm is evaluated based on the order of growth (e.g. logarithmic, linear, quadratic, etc) of the worst-case time complexity with respect to the increasing size of the input. The worst-case running time of a linear time complexity algorithm will increase linearly with the sample size. Often algorithms that rely on `between member' operations within the sample (such as a classic Self-organising Map Algorithm) will have running times that scale quadratically with increasing sample size or worse.} More recently complexity has been applied by \citet{bartlett2022assessing} as a new approach for exoplanet characterization  with potential applications to biosignature detection. 

In this paper we apply the coarse-grained complexity, as calibrated in S19 and without re-training, to a much larger data set from the Pilot Survey of the Evolutionary Map of the Universe (EMU-PS, \citet{2021PASA...38...46N}). While the primary goal of the EMU Pilot Survey is to test and refine observing parameters and strategy for the main survey, the EMU-PS in itself presents opportunity for new discoveries. Recently self-organising maps have been used successfully to detect unusual sources with reference to the broader sample (ensemble) using a sub-set of components from complex sources selected from EMU-PS catalogue data \citep{Nikhel}. The method used in the present paper is based on a complexity measure that has the advantage of being efficient to compute as it does not require computing pair-wise distances between observations within the broader sample 
and does not require, hence is not confined to, traditional source extraction or existing catalogue data. We partition the data using frames rather than identified sources, computing the course-grained complexity within a sliding frame (square aperture) across the image. An important feature of the scan method used is that the frames are sampled from the EMU-PS data in a blind manner (that is without using a source extraction tool or existing catalogue data). This helps reduce the risk of producing a sample that is biased towards preconceived notions of what is interesting, referred to as expectation bias by \citet{norris2017} and \citet{Robinson87}. The approach can also be used without the prior identification of complex sources. This is intended to assist with the identification of the unexpected in new and large data with the goal of new scientific discoveries and surprise. 

We evaluate the effectiveness of the approach at finding new and unusual objects by using a Zooinverse project to produce crowd-sourced (from amongst astrophysicists) labels for frames produced by the scan. These labels can be used to generate a partition boundary for anomalies which can be used to create an anomaly catalogue. An effective anomaly partition is an intended product from this work, providing: a concentrated search space, rich in unusual or anomalous objects, intended to provide an efficient tool for assisting with scientific analysis and new discoveries. While not every object in the partitioned space (or anomaly catalogue) will necessarily be truly `anomalous', the aim is to define  a partitioned space (anomaly catalogue) where almost all contained frames will be interesting in some fashion (i.e. complex).

This paper proceeds as follows:  in section \ref{sec:emupilotdata} we introduce the EMU-PS data as a test case for using the coarse-grained complexity measure to identify anomalous objects in future large-scale surveys. The EMU-PS data is likely to contain many rare, unusual and anomalous objects, providing a better representation of the complexity tail of still much larger future surveys (compared to the S19 analysis). 

In section \ref{sec:appcomplexity} we describe the theoretical foundations for the coarse-grain complexity measure and show how it can be practically computed. This section also outlines the implemented methods, including the scanning method applied to the EMU-PS, and details the use of a Zooniverse project to crowd-source labels for EMU-PS frames. The labels are used to evaluate the effectiveness of coarse-grained complexity partitions for identifying complex and unusual objects.

Section \ref{sec:complexityimages} commences by showcasing a variety of frames captured in the complexity tail containing complex structures and unusual objects (with additional examples and a discussion of their characteristics provided in Appendix \ref{appendix:examples}). We then provide an overview of zoo results used to produce consensus truth labels for EMU-PS frames. We conclude this section by providing an evaluation of the effectiveness of alternative partition boundaries. 

Section \ref{sec:Conclusions} details the application of these boundaries in constructing anomaly catalogues and provides  a discussion of potential classification errors and considerations. Section \ref{sec:summary} concludes with a summary of the key outcomes and conclusions drawn from this work. In Appendix \ref{appendix:examples} we provide detailed descriptions of 36 example objects found in high complexity frames.



\section{EMU Pilot Survey Data}
\label{sec:emupilotdata}

The Pilot Survey of the Evolutionary Map of the Universe (EMU-PS) was observed at 944MHz using the Australian Square Kilometre Array Pathfinder (ASKAP) telescope. The ASKAP telescope consists of 36 12-metre antennas spread over a region 6-km in diameter at the Murchison Radio Astronomy Observatory in Western Australia. EMU-PS covers 270 square degrees of an area covered by the Dark Energy Survey at a spatial resolution of $\sim$ 11–13 arcsec \citep{2021PASA...38...46N}. 

While the primary goal of the EMU Pilot Survey is to test and refine observing parameters and the strategy for the main survey, the pilot in itself presents opportunity for new discoveries. Experience has shown \citep{norris2017} that whenever we observe the sky to a significantly greater sensitivity, or explore a significantly new volume of observational phase space, we make new discoveries. This goal has already been demonstrated through the successful identification of a new class of radio object, odd radio circles (ORCs, \citet{norris2021a,10.1093/mnrasl/slab041, norris22}, Figure \ref{fig:examples} and \ref{fig:orcs}). 

The observations and data reduction for EMU-PS are fully described
by \citet{2021PASA...38...46N} so here we restrict our description to the data product used in the present paper.
The data were taken in 10 overlapping tiles, each covering an area of 30 square degrees. These were then merged while correcting for the primary beam response, to produce a single image covering 270 square degrees.  Here we use the
``native'' resolution product (i.e. not convolved to a common beamsize), giving a synthesised beam of about
11 $\times$ 13 arcsec with an rms sensitivity of about 25 $\mu$Jy/beam. 
Source extraction using the {\it Selavy} tool \citep{whiting17} found a total of  220,102 radio components, of which 178,821 are ``simple sources'', which are either unresolved point sources or can be fitted by a single Gaussian. The remaining 41,181 ``complex'' sources  range from small extended sources to giant radio galaxies, and include a number of objects with complex morphology which are the sources of primary interest in the present paper. However it is important to note that, to avoid bias, the {\it Selavy} extractions are not used at all in this paper, which instead works directly with the image data.

The EMU-PS data is likely to contain examples of anomalous and unexpected objects, providing a better representation of the complexity tail of still much larger future surveys such as EMU (compared to previous experiments using ATLAS data in S19). It is in this tail that future discoveries are likely to be made.


\section{Methods}
\label{sec:appcomplexity}
In this section we describe the theoretical foundations for the coarse-grained complexity measure and show how it can be practically computed. We also outline the practical implementation of this approach and the method used to scan the EMU Pilot Survey data.

We describe how we used a Zooniverse project to crowd-source labels of EMU-PS frames to evaluate the coarse-grain complexity as a tool for partitioning, and hence identifying, anomalous objects. We then discuss the methods and challenges involved in sub-sampling for the zoo from the very large number of frames produced by the EMU-PS scan and conclude with details about how complexity partitions will be evaluated using the labelled data.


\subsection{Coarse-grained complexity}

The coarse-grained complexity as defined in S19 is based on the notions of \textit{effective complexity} \citep{GellMann94, gell1996information} and  \textit{apparent complexity} \citep{Carroll}. The apparent complexity is a measure of the entropy $H$ of an object $x$ computed after applying a smoothing function $f$, expressed as $H (f (x))$. The Shannon entropy of a probability distribution P can be defined as the expected number of random bits that are required to produce a sample from that distribution:
\begin{equation}H(P) = -\sum_{x \in X} P(x)\log P(x) \,. \end{equation}

By Shannon's Noiseless Coding Theorem the minimum average description length $L$ of a sample is close to the Shannon entropy:
\begin{equation}
H(P) \leq L \leq H(P) + 1 \,. 
\end{equation}
The Kolmogorov complexity $K (f (x))$  can be used as a proxy for the entropy of the smoothed function $H (f (x))$, as proposed by \citet{Carroll}. The analogy between the concept of entropy and program size has been previously recognised \citep{Chaitin}. The Kolmogorov complexity of $x$ is the length of the shortest binary program $l(p)$, for the reference universal prefix Turing machine $U$, that outputs $x$; it is denoted as $K(x)$:
\begin{equation}
K(x) = {\rm min}_{p}\{l(p):U(p)=x\} \,. 
\end{equation}

A thorough treatment is provided by \citet{Li}. The Kolmogorov complexity has the advantage of being well-defined for a particular description of a system such as an image of a galaxy. This is not the case for the Shannon entropy which is defined in terms of the possible states of the system. While the Kolmogorov complexity is uncomputable, its upper bound can be reasonably approximated by the compressed file size $C (f (x))$ using a standard compression program \citep{Carroll}, such as \texttt{gzip}. 

The issue with using the approximated Kolmogorov complexity directly as a measure of complexity is that it is maximised by random information. Intuitively a complexity measure should provide low values for random data that does not contain structure that is of interest to the observer \citep{Zenil}. \citet{Carroll} have shown that the apparent complexity measure is able to achieve this by applying a smoothing function $f$ to the input $x$ (which removes fine-grained noise while preserving the coarse-grained structure of the image). While the Kolmogorov complexity of a random sequence is large, the apparent complexity of the same sequence becomes small with smoothing, as fluctuations are removed where the average or median information content becomes homogeneous at the coarse-grained resolution. Accordingly we define the coarse-grained complexity, like the apparent complexity, as the compressed description length of regularities and structure after discarding all that is incidental. The coarse-grained complexity will be small for both simple and random sequences. The coarse-grained complexity measure extends the idea of the apparent complexity by incorporating calibration of the measurement resolution in alignment with expert distinctions between meaningful structure and noise. This is achieved by adjusting the the measurement resolution through a smoothing function so that complexity values correctly partition data that has been expertly labelled  (i.e. by human astronomers).

The objective of applying the smoothing function $f$ when deriving $C(f(x))$ is to remove incidental or random information, such as instrumental noise, that have no regularities comprehensible to the observer even though it may have a physical basis. Comprehensibility here is defined with respect to the observer of the information, in this case scientists with specific interests. Comprehensible information has a structure within a feature space, which in the case of images refers to the spatial distribution of bits of information across available channels. 

Importantly, the coarse-grained complexity measure of an image does not rely on the presence of any particular structures or structural elements and desirably does not demonstrate sensitivity under affine transformations based on the implementation in S19. The calibration of the measure makes only explicit assumptions regarding the choice of the coarse-graining level and the scale of the image. Previous data are used only to calibrate the coarse-graining level (i.e. the appropriate measurement resolution) and only a small sample of the relevant data type is required.

Coarse-grained complexity runs into obstacles as a well-defined measure of complexity. Firstly, the uncomputability of the Kolmogorov complexity prohibits the concept from being defined in terms of an optimal compression. It has been proven by \citet{Chaitin1} that there can be no procedure for finding all theorems that would allow for further compression. Furthermore the problem of distinguishing between meaningful structure and incidental information, especially in finite data, may fail to be well-defined. Different smoothing functions and different coarse-graining levels will retain different distinct regularities in the data.

These theoretical challenges in objectively defining the coarse-grained complexity can be circumvented when the approach is applied to the segmentation of observations by complexity. Here the coarse-grained complexity can be calibrated to coincide with notions of complexity adopted by the observer (i.e. expert astronomers) and evaluated using expertly labelled data.

\subsection{Scanning the EMU Pilot Survey}
\label{sec:complexity_calc}

\begin{figure}
\includegraphics[width=8cm]{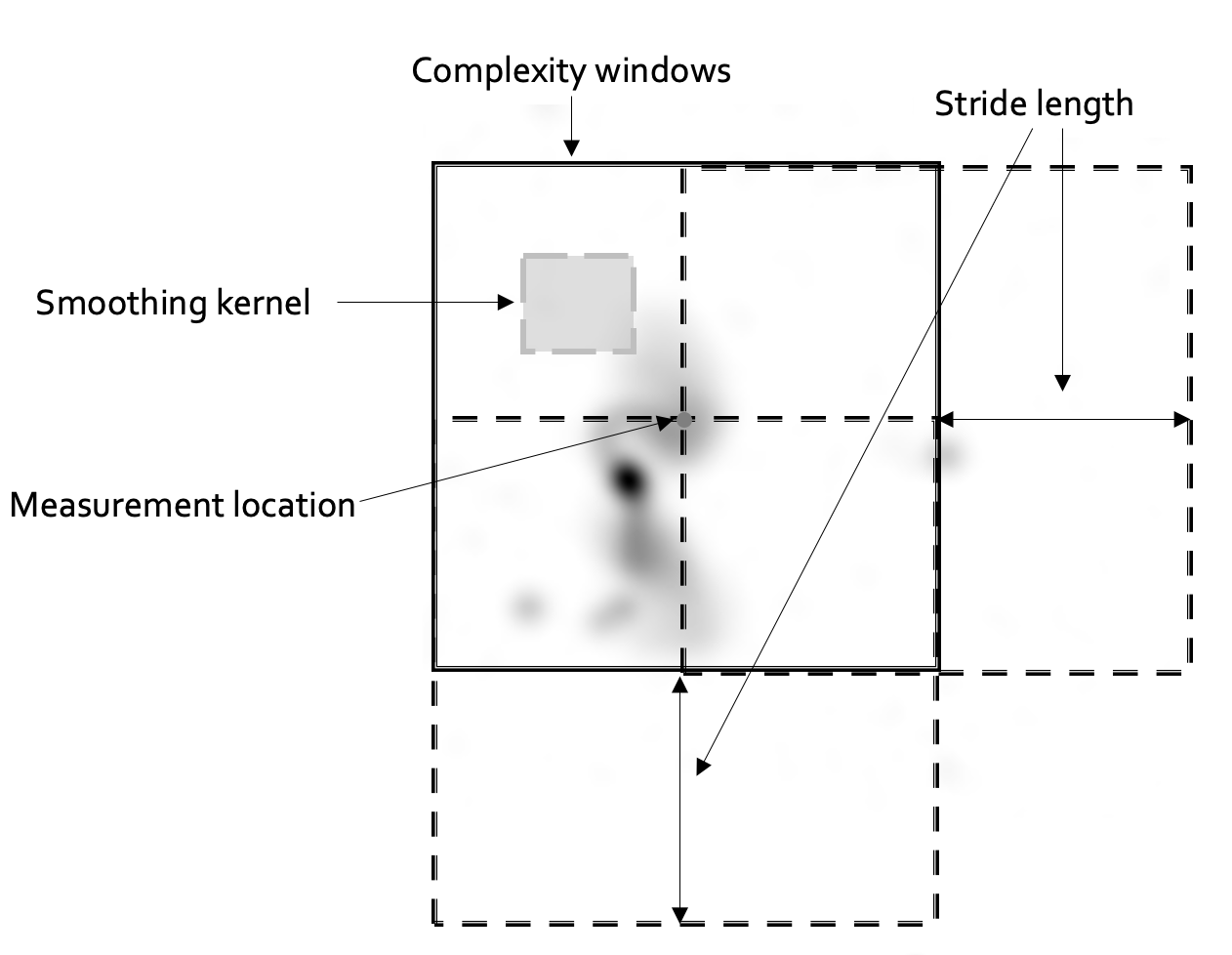}
\caption{\label{fig:stride} The definition of frame size (side length of solid black square), stride length (offset between position of solid black square and dashed black squares) and smoothing kernel size (grey shaded rectangle). The size of the complexity frame is chosen such that most (but not all) extended sources will be fully contained. The stride length is chosen to be one quarter of the side length of the frame, however the illustration above depicts one half for illustrative purposes only. These are the only free parameters used in this scanning method.}
\end{figure}

We perform a scan of the EMU Pilot Survey data. When performing the scan we estimate the coarse-grained complexity within a sliding frame of fixed size, rather than working with pre-selected sources. The coarse-grained complexity is computed within each frame using the approach implemented in S19 without re-calibration. The approach is based on \texttt{gzip} compression using Lempel-Ziv (LZ77) and Huffman coding. 
Each frame is a \textit{blind} sample from the EMU-PS data, with a high likelihood of many containing no sources, or only part of the structure of a large complex source. After computing the complexity in a frame, it is shifted based on a defined stride length to the right and computed again, progressing in this manner until it overlaps with the edge of the image. Once the frame overlaps with edge of the image it returns to the starting column and is shifted down by the defined stride length, progressing in this manner until the frame overlaps with the lower edge of the image. Figure \ref{fig:stride} illustrates the sliding frame and the associated parameters. The choice of frame size, stride length and smoothing kernel size are the only free parameters used in this method. 
For the EMU-PS image, a frame is defined to be a 256 $\times$ 256 pixel region (equivalent to a span of approximately $\sim 12$ arcmin) that we slide based on a stride length of 64 pixels. This frame size exceeds the angular size of most known radio sources in the EMU-PS field, with the exception of a few Giant Radio Galaxies such as those shown in Figures 16 and 28 of \citet{2021PASA...38...46N}. 

An important feature of the scan method is that the frames are \textit{blind} samples from the EMU-PS data. This helps reduce the risk of biasing the sample to preconceived notions of what is complex and interesting. It also avoids restricting the sampling to only regions of an image that are already represented in existing catalogue data. As a consequence of this method, many frames will not contain any detectable sources or objects of interest. Conversely, some frames may contain a part of a source but not the entire object. To help minimise this risk, a stride length of one quarter the span of the frame size was selected. The overlapping frames provide better coverage of the EMU-PS data for computing complexity, as they improve the probability of capturing entire complex structures, or interesting parts of structures in a single frame.

The smoothing kernel size defines the measurement resolution for the coarse-grained complexity scan. Smoothing is implemented using a median filter $f$ applied to frames from the EMU-PS image $x$. The smoothing kernel size $h$, is in this case calibrated to 10 pixels, consistent with the learned smoothing kernel size from S19. This allows the generalisability of the calibration adopted in S19 to be evaluated using new data from the EMU-PS. The kernel size was calibrated in S19 using ATLAS DR1 data using sources that had been manually classified as simple and complex. The kernel size was chosen by maximising the difference between the average coarse-grained complexity of observations labelled complex and simple thus coinciding with notions of complexity of interest to the observer (in this case expert astronomers). A median filter was chosen as the filter type because it completely removes noise and incidental values in regions that are predominately without flux measurement (i.e. where no sources are detected) and retains the strength of signals in regions dominated by actual measurements. The median filter is also effective at preserving edges (compared to, for example, a Gaussian filter), given the expected noise in each frame. 

This approach does not control for the distance of objects contained within each frame and accordingly the measurement resolution does not vary depending on the observed scale of sources contained within each frame. Rather this approach provides a consistent evaluation of complexity for each section of sky covered by the sliding window, irrespective of the sources it contains.

The approach is implemented in accordance with the procedure followed in S19 to calculate the coarse-grained complexity within each frame \footnote{Unlike the procedure followed in S19, frames are not cropped. Instead the complexity is computed for the entire 256 $\times$ 256 pixel frame.}. The same parameters were applied consistently across the entire EMU-PS radio mosaic. Consistent with S19, a pixel intensity threshold is set at the 90th percentile whereby all values below the threshold are set to zero. S19 explores the sensitivity of the method to the threshold value, showing the effectiveness of the threshold selected compared to alternatives. To estimate $C (f (x))$ as an upper bound on the Kolmogorov complexity $K (f (x))$ we calculate the size in bytes of the \texttt{gzip} compressed image. The signal-to-noise ratio (SNR) was calculated using the reciprocal of the coefficient of variation for the smoothed array as was implemented in S19. Our analysis in S19 found that a noise measure complemented the complexity measures and that the the noise and complexity plane improved results when identifying complex and unusual sources.

\subsection{Crowd-sourced evaluation of frames}

\begin{figure*}
\includegraphics[width=16cm]{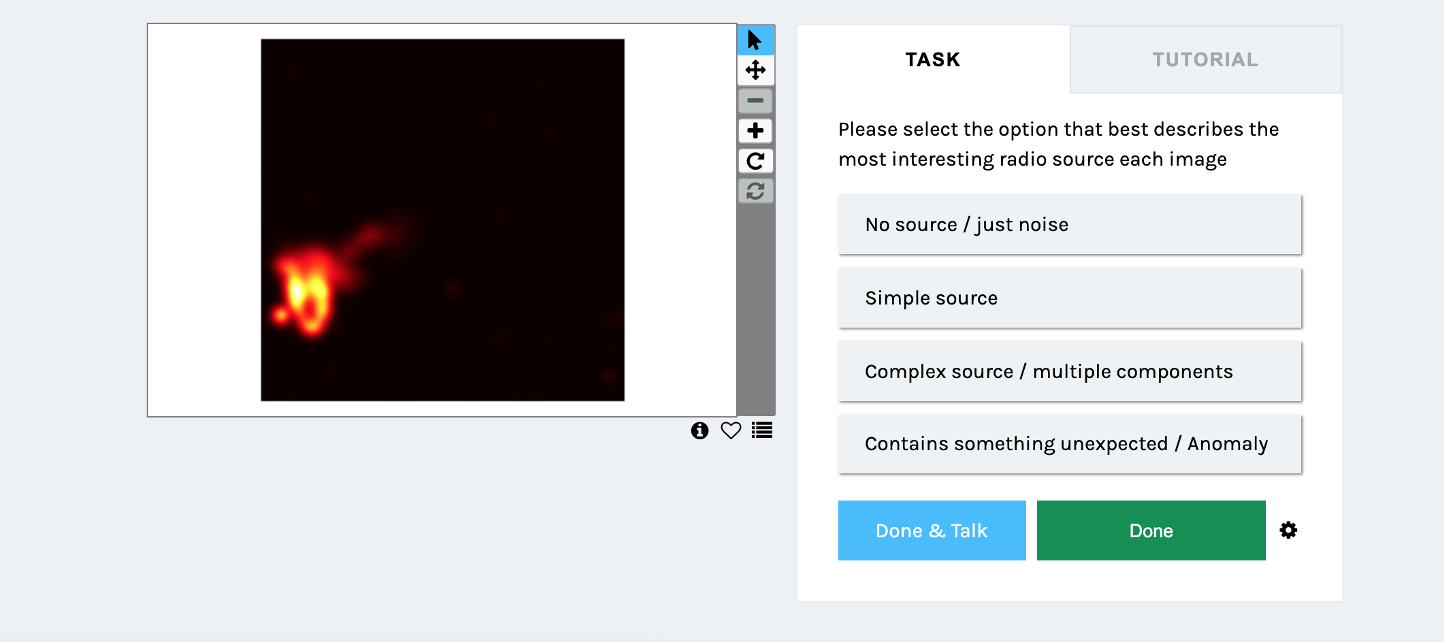}
\caption{\label{fig:wflow} Example workflow from Anomaly in the EMU Zoo. In this example we pick a frame containing 2MASX J21291901-5053040, bent tail radio galaxy near ABELL 3771 cluster.}
\end{figure*}

To identify anomalous observations the sample can be a partitioned by determining an appropriate threshold, that is a complexity value in the tail of the complexity distribution, above which defines an anomaly. A threshold drawn at a low complexity value will produce a very large sample of potential anomalies. If the goal is to identify complex and unusual objects while minimising the search space for new discoveries (to improve efficiency), then the objective becomes to partition at the largest complexity value that is still exceeded by as many of the most interesting objects as possible. 

In S19 the sample was segmented based on the coarse-grained complexity and the signal-to-noise using unsupervised clustering methods (Gaussian Mixture Models), and the results were evaluated using truth labels based on expert classification of the ATLAS data. In the case of the EMU-PS scan we have a much larger sample size of blindly sampled frames, not necessarily containing sources, for which truth labels do not exist. 

Truth labels are required to evaluate the effectiveness of alternative complexity thresholds for partitioning anomalous sources. To provide truth labels for the frames produced by the complexity scan we ran a project on the \url{Zooniverse.org} platform, titled \textit{``Anomaly in the EMU Zoo''} (hereafter zoo), requesting expert astronomers to evaluate an unbiased sample of frames sub-sampled from the EMU-PS scan. Consensus from the zoo labels was then used to evaluate the Recall, Informedness and Precision associated with prospective partition boundaries. Recall, Informedness and Precision are explained in detail in appendix \ref{appendix:evaluation}.

Expert volunteers were approached from within the Evolutionary Map of the Universe Survey Project and at the SPARCS 2021 conference. A sub-sample of 1627 frames 
from the EMU-PS scan ($n_{\mathrm{total}}$=365,000) were presented to volunteers for classification through the Zooniverse project. 44 volunteers participated in the project, with 10 of these classifying more than 500 frames. 

The zoo asked the expert volunteers to evaluate frames sub-sampled from the EMU-PS Scan and to select an option that best describes the most interesting radio sources in each frame before moving on to the next. Sub-sampling is discussed in section \ref{sec:Sub-sampling}. An example of this workflow is shown in figure \ref{fig:wflow}. The four options presented for selection were:
\begin{itemize} 
\item No sources/just noise
\item One or more simple sources/unrelated simple sources
\item At least one complex source/sources with multiple components
\item Contains something unexpected/Anomaly
\end{itemize}

The zoo distinguished between complex and extended sources with multiple components, and sources that were deemed by the volunteers to be truly unexpected or anomalous. This distinction between complex and anomalous sources enabled the evaluation of a complexity threshold that could be used to partition a smaller sample of interesting frames that had high Recall and Informedness with respect to anomalies. Ensuring that the partitioned data has high Recall and Informedness with respect to anomalies only supports science objectives for studying and identifying interesting objects that may result in new discoveries and minimises the search space for such objects. Retaining more typical complex objects to maintain a high  Recall was not seen as an issue, given the potential overlap in the complexity values of more typical complex objects and truly anomalous objects, so long as the search space remained small as measured by the false positive rate. This also helps account for the subjective nature of assigning truth labels, where only some volunteers may assign labels based on unexpected or subtle unusual features belonging to what otherwise would appear a more typical complex object.

Only frames converging on a label through majority consensus were used to evaluate the effectiveness of the complexity for identifying anomalous sources. Those frames from the zoo sample where majority consensus was not reached were excluded from the evaluation.

\subsection{Sub-sampling of EMU-PS frames} 
\label{sec:Sub-sampling}

The EMU-PS data covers an area containing approximately 220,000 catalogue sources \citep{2021PASA...38...46N} and can be used to generate approximately 365,000 sampled frames based on the scan parameters selected. This makes evaluating potential partitions more challenging, as it is not feasible to have experts inspect and classify the hundreds of thousands of frames produced by the scan.

Sub-sampling can be used to make the evaluation of alternative complexity partitions in the EMU-PS data more feasible. Here the data size is reduced by selecting a subset from the original sample. Expert evaluation of this subset is more feasible on shorter time scales.

Selecting sample frames from the EMU-PS scan to be evaluated by the zoo was done in two phases. The first phase involved blind unbiased sub-sampling (n=1528) by sampling proportionally across the distribution of complexity values. Sampling was performed on non-overlapping frames (using the first occurrence) without replacement. The second phase involved enrichment sampling (n=99) from a EMU-PS scan done at a stride length of 32 pixels, half the original stride length. The reduced stride length samples from the EMU-PS more thoroughly, and so can capture frames of higher complexity than the scan done at a stride length of 64 pixels, extending the tail of high-complexity values.  Due to the much larger number of frames produced (n = 1.4 million) when using the smaller stride length, only frames above the 99.5th percentile complexity value were retained (n = 7,000) and  sub-sampling was performed uniformly across these values to produce the enrichment sample (n=99). Sub-sampling was performed inward from the tails of the complexity distribution of samples produced from the complete scan, leaving out frames close to the median complexity value. Sub-sampling from within the this range was not deemed as necessary as discussed in detail in Appendix \ref{appendix:lefout}.

The purpose of the enrichment sample was to supplement the tail of the 64 pixel stride sub-sample distribution with frames of complexity above the 99.5th percentile. The sample size for the zoo was limited (n=1627) to ensure every frame could be evaluated with sufficient multiplicity to achieve consensus, however this results in poor sub-sampling from the far right (high complexity) tail. The enrichment sample was intended to provide better representation of the type of observations found within frames beyond the 99.5th percentile. 

\subsection{Measuring the effectiveness of partitions}

The truth labels derived from the zoo were used to evaluate the effectiveness of alternative partition boundaries for identifying anomalous sources. We evaluated alternative partition boundaries using a binary classification approach, evaluating both the Recall, Informedness and Precision as discussed further in Appendix \ref{appendix:evaluation}, and illustrated in Table \ref{tab:confusionmatrix}. 

The objects in the zoo that were selected in the first phase of sampling (without the enrichment sample added) allow the Recall to be assessed at a given complexity threshold. 
The enrichment sample creates a bias for the purpose of calculating Recall by over-representing high complexity objects in the total zoo sample and accordingly was not included in the non-bias sample used to evaluate prospective partitions. The second phase, where the enrichment sample is added, was performed to increase the sample size of unusual observations and provide further analysis of the Precision within the complexity tail. 

When evaluating Recall and Precision we consider the positive class to include only frames containing something unexpected or anomalous and the negative class to include all other frames. Limiting the positive class to only frames containing unexpected or anomalous sources provides a measurement framework that will assist in defining an anomaly catalogue with the objective of supporting new and novel scientific discoveries in an efficient manner. A catalogue containing all extended sources would be much larger and would increase the search space for novel and anomalous objects, making the discovery process less efficient.

We note that many objects with complex and interesting morphology will have familiar features and accordingly not be considered anomalies. As science progresses and new discoveries become familiar the frequency of finding such familiar objects at high complexity will increase. Misclassification of such complex sources as anomalies is not deemed an issue so long as the overall search space (or catalogue size) defined by the complexity measure remains small enough to support an efficient search for novel discoveries. This can be achieved by evaluating the trade-off between a high Recall and a high false positive rate through maximising the Informedness (see Appendix \ref{appendix:evaluation}).

A high Informedness value can still be accompanied by low Precision, as we expect in this case, due to the the low frequency of sources deemed by consensus as being anomalous compared to those being deemed complex (see Table \ref{tab:zooclassification}) and the potential overlap in complexity values of familiar complex objects and anomalous complex objects. To determine if low Precision is explained by the misclassification of familiar complex objects we also evaluate the Precision associated with prospective complexity partitions by defining the positive class to include both anomalies as well as frames containing at least one complex source or a source with multiple components. These sources will still be interesting to some observers and should not hinder the search for novel discoveries so long as the false positive rate and overall search space remains small.

\section{Results}
\label{sec:complexityimages}

\subsection{EMU-PS scan results }

\begin{figure}
\includegraphics[width=8cm]{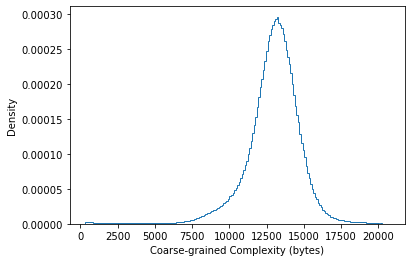} 
\caption{\label{fig:sample_clean} Complexity distribution for EMU-PS data data after removal of low complexity background data (frames with the minimum complexity value). The left tail contains frames without sources or point sources, predominately point sources toward the centre. }
\end{figure}

\begin{figure*}
\includegraphics[width=16cm]{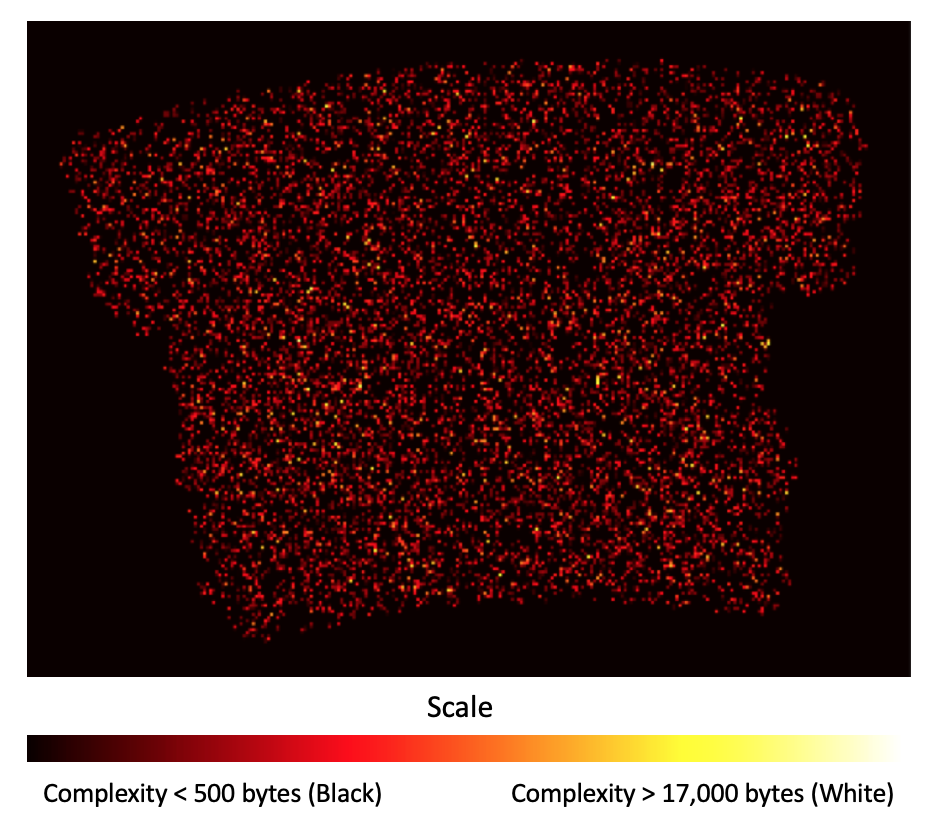}
\caption{\label{fig:hmap} A complexity `heat map' of the EMU-PS region, where the complexity of the frame is indicated by the brightness of the pixel. While many of the frames are red coloured, indicating the presence of a simple source, there are a few yellow or white pixels, indicating more complex sources or low-surface brightness structures. The most complex frames seem to be randomly distributed across the survey region, with the exception of the edges, where incidental structure increases the complexity. The region outside of the EMU-PS is coloured black.}
\end{figure*}

\begin{figure}
\includegraphics[width=8cm]{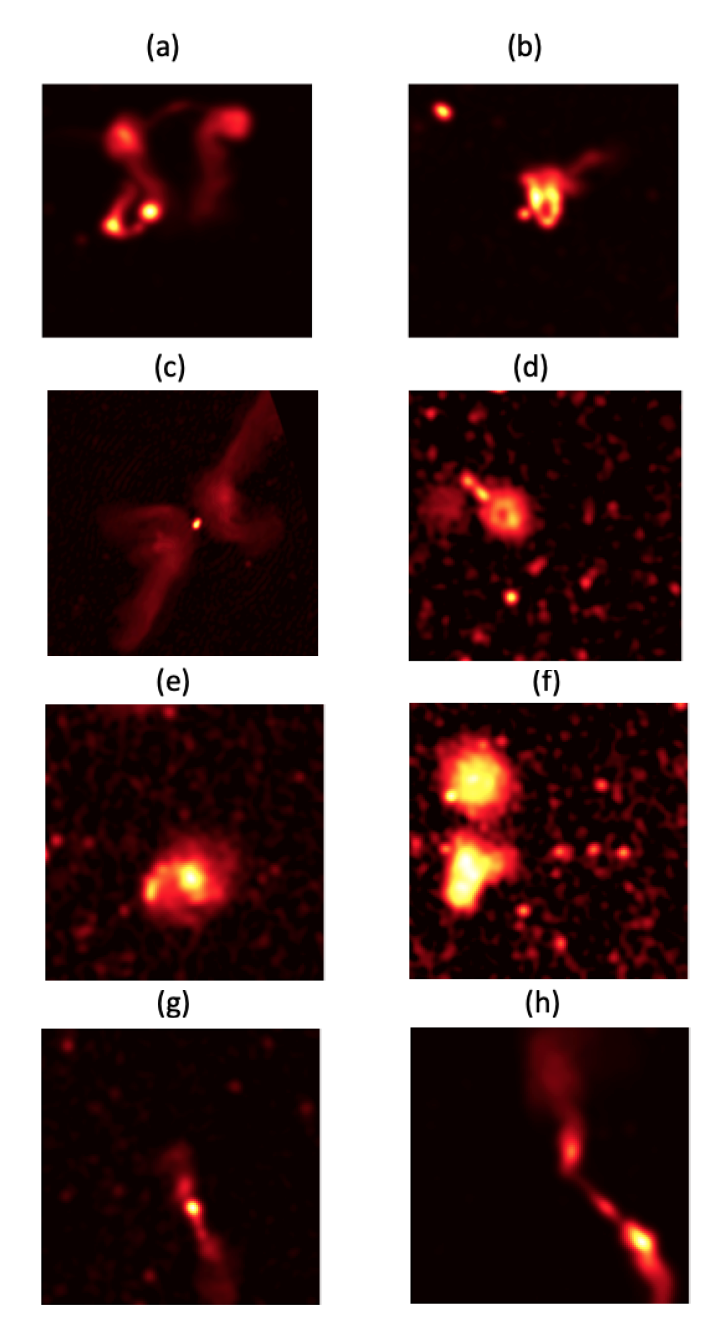}
\caption{\label{fig:examples} Examples of objects found within frames sampled above the 99th percentile complexity value from the EMU-PS Complexity Scan. These examples illustrate the breadth of objects found within frames in the complexity tail including, (a) the unusual source PKS 2130-538 \citep{Otrupcek} known as the Dancing Ghosts \citep{2021PASA...38...46N}, (b) a bright wide angle tail source on 2MASX J21291901-5053040 in cluster Abell 3771, (c) the large X-shaped radio source PKS 2014-55 (2MASX J20180125-5539312), (d) two odd radio circles, EMU PD J205842.8–573658 (ORC2) and EMU PD J205856.0-573655 (ORC3), (e) a face-on spiral galaxy NGC 7125, (f) 2MASX J20483764-4911157  FR-II remnant, (g) DES J202818.12-492408.4 FR-I potential Double-double radio galaxy and (h) FR-I extended radio source with host galaxy 2MASX J21512991-5520124.
}
\end{figure}

\begin{figure}
\includegraphics[width=8cm]{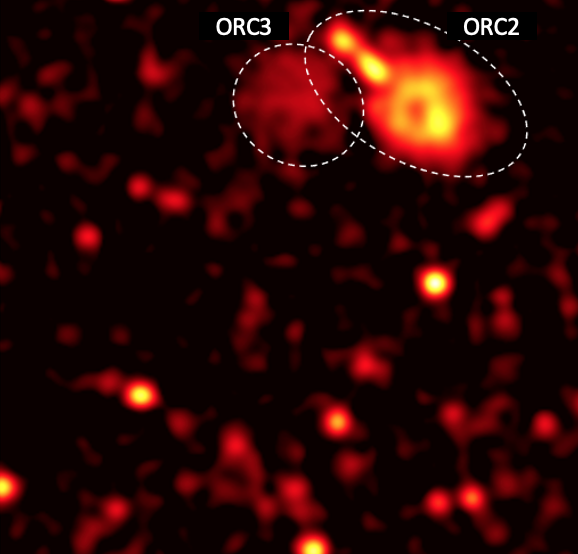}
\caption{\label{fig:orcs} An example frame with a complexity value above the 99.5th percentile (> 17,000 bytes). In this frame EMU PD J205842.8–573658 (labelled ORC2) in the top right is the dominant source. This frame also contains EMU PD J205856.0-573655 (labelled ORC3).}
\end{figure}

We now apply our method to analyse radio images. We scan the EMU-PS data, examine the distribution of complexity values for each frame, and showcase a variety of complex and unusual objects captured in the complexity tail. 

We performed a complexity scan of the  EMU-PS data using the method described above. The distribution of complexity values from each sampled frame is shown in figure \ref{fig:sample_clean}. Complexity values are shown in figure \ref{fig:hmap} as a heat map over the EMU-PS field.

Visual inspection of a sub-sample of frames shows that the high-complexity value tail of the distribution comprises unusual, complex and extended objects. Extended sources featured heavily in the tail above the 95th percentile ($\sim15,000$ bytes), with wide angle tail galaxies and objects of a more anomalous appearance apparent from above the 99th percentile ($\sim16,500$ bytes). Findings demonstrate the effectiveness of this approach in recovering known extended and complex structures. Examples of these objects, sampled above the 99th percentile, are shown in figure \ref{fig:examples}. 
These examples illustrate the breadth of objects found within frames in the high-complexity value tail, including, (a) the unusual source PKS 2130-538 \citep{Otrupcek} known as the Dancing Ghosts \citep{2021PASA...38...46N} (b) a bright wide angle tail source on 2MASX J21291901-5053040 in cluster Abell 3771, (c) the large X-shaped radio source PKS 2014-55 (2MASX J20180125-5539312), (d) two odd radio circles, EMU PD J205842.8–573658 (ORC2) and EMU PD J205856.0-573655 (ORC3), (e) a face-on spiral galaxy NGC 7125, (f) 2MASX J20483764-4911157 an FR-II remnant, (g) DES J202818.12-492408.4 an FR-I potential double-double radio galaxy and (h) an FR-I extended radio source with host galaxy 2MASX J21512991-5520124.\footnote{These objects, with some additional examples, are shown in Appendix \ref{appendix:examples}, where we discuss the more unusual features of the objects, grouping these examples together in the currently most commonly used radio-morphological classification schemes. These include a previously unreported Giant Radio Galaxy with Largest Linear Scale (LLS) of 1.94 Mpc shown in panel 5D of figure \ref{fig:complexexamples}.} 

The EMU-PS has already produced valuable scientific discoveries, including the identification of a new class of radio object called Odd Radio Circles (ORCs, \citet{norris2021a, 10.1093/mnrasl/slab041, norris22}). Examples, labelled ORC 2 and ORC 3, are shown in the high complexity frames presented in Appendix \ref{appendix:examples}, figure \ref{fig:examples} (d) and figure \ref{fig:orcs}. ORCs provide examples of recent discoveries coming from EMU-PS that are found in the far right tail of the complexity distribution. The ability of the course-grained complexity measure to segment these sources above a high complexity value in the distribution tail, in this case above the 99.5th percentile, supports the effectiveness of this approach at identifying scientifically interesting observations in large data in an efficient manner. Being able to capture and segment complex structures and unusual observations in the far right tail provides a smaller and more concentrated search space that can potentially improve the speed and efficiency of making new discoveries. As discussed in S19, the worst-case linear time complexity of the method also makes it computationally efficient to implement.

\subsection{Zoo results }

We use truth labels determined from the zoo to evaluate the effectiveness of alternative partitions for identifying anomalous objects. As an immediate benefit, an effective partition can be used to identify frames containing complex structures and unusual objects from the EMU-PS data and build an anomaly catalogue. An effective boundary can also be used when analysing new, and even larger surveys, for interesting objects and new discoveries. As an example, the thresholds determined from the EMU-PS data can be used to identify the most interesting frames in the subsequent full EMU survey which is anticipated to capture approximately 40 million sources.

The EMU-PS scan contains approximately 365,000 sampled frames making evaluation of all frames from the zoo infeasible. Through sub-sampling of the EMU-PS Scan, the sample size for the zoo was limited (n=1627) to ensure all frames could be evaluated by a larger number of volunteers. However this results in poor sub-sampling from the far right (high complexity) tail. The enrichment sample was intended to provide better representation of the type of observations found within frames beyond the 99.5th percentile. Classification counts for each of the zoo classes are shown in table \ref{tab:zooclassification}. This table includes the anomaly counts both before and after the inclusion of the enrichment sample.

\begin{table}
\begin{tabular}{l c c c}
\hline \hline
Classification & Count:  & Count: & Count: \\
 & Non-bias sample & Enrichment & Combined Sample \\
 & (n=1528) & (n=99) & (n=1627)  \\
\hline 
Anomalous& 7 & 17 & 24\\
Complex& 366 & 57 &423 \\ 
Simple& 1059 & 12 &1071 \\ 
No source& 31 & 0 &31 \\ 
No consensus& 65 & 13 &78 \\ 
\hline
\end{tabular}
\caption{\label{tab:zooclassification} Classification counts by consensus for each of the options provided in the EMU-PS Anomaly in the EMU Zoo. The enrichment sample supplements the zoo sub-sample with frames measuring complexity above the 99.5th percentile. This creates a bias for the purpose of calculating Recall but improves the sample size for calculation of Precision above prospective partition thresholds. This data shows that 86\% of the enrichment sample, where consensus was reached, contains frames classified as containing complex and anomalous sources. }
\end{table}

Only frames converging on a label through majority consensus were used to evaluate the effectiveness of the complexity to identify anomalous sources. Those frames from the zoo sample where majority consensus was not reached were excluded from the evaluation.

Table \ref{tab:condist} shows that 99.8\% of the zoo sample frames received evaluations from 3 or more volunteers and 94\% from 5 or more. We used the criterion  that only frames receiving 5 or more evaluations were used to evaluate consensus, and be given a reliable `truth' label. These restrictions were imposed to avoid the results being impacted by outlier evaluations that differed from the majority of expert volunteers.

The average number of evaluations per frame was 6.85, with a relatively even number of average evaluations across the outcome labels as shown in Table \ref{tab:conclass}. Table \ref{tab:conclass} also shows that where an outcome label was assigned, the average consensus level was 70\% or greater. The number of zoo sample frames converging on a majority consensus in each class is shown in table \ref{tab:zooclassification}.

\begin{table}
\begin{tabular}{l l}
\hline \hline
Number of Evaluations & Percentage of samples \\
\hline 
1 & 0.0\%\\
2 & 0.2\%\\
3 & 0.9\%\\
4 & 5.2\%\\
5 & 12.4\%\\
6 & 23.7\%\\
7 & 24.3\%\\
8 & 18.6\%\\
9 & 10.5\%\\
10 & 3.5\%\\
11 & 0.6\%\\
12 & 0.1\%\\
\hline
\end{tabular}
\caption{\label{tab:condist} Distribution of evaluation counts across the zoo sample}
\end{table}

\begin{table}
\begin{tabular}{ l  c  c }
\hline \hline
Zoo label & Average number of votes & Average consensus \\
\hline 
Anomalous& 6.8 & 70\%\\
Complex& 6.9 & 77\% \\ 
Simple& 6.9 & 87\% \\ 
No source& 6.5 & 78\% \\ 
No consensus& 6.4 & 48\% \\ 
\hline
\end{tabular}
\caption{\label{tab:conclass} Average number of evaluations and consensus level across the outcome labels}
\end{table}

\subsection{Defining Complexity boundaries}

\begin{table*}
\centering
\begin{tabular}{l c c c c}
\hline \hline 
 \multicolumn{5}{c}{ EMU-PS Anomaly in the EMU Zoo sub-sample excluding enrichment sample and no consensus zoo labels (n=1463)} \\
\hline 
Partition boundary & Recall & False Positive Rate & Informedness  & Precision \\
 & (Anomalous only) & (Anomalous only) & (Anomalous only) &  (Anomalous only) \\
\hline 
Complexity $\ge$ 14,000 bytes  & 1.00 & 0.37 & 0.63 & 0.01 \\
Complexity $\ge$ 14,000 bytes, SNR $\ge$ 0.14 & 1.00 & 0.17 & 0.83 & 0.03 \\
\hline 
\end{tabular}
\caption{\label{tab:miscresults} Informedness and Precision measures at various partition boundaries. These measures are explained in detail in appendix \ref{appendix:evaluation}}
\end{table*}

\begin{table}
\begin{tabular}{c c c c}
\hline \hline
Enriched sample & Prediction:  & Prediction: & Total\\
Zoo Label & Anomaly & Other  \\
\hline 
Anomalous& 23 & 1 & 24\\
Complex& 58 & 365 & 423  \\ 
Simple& 22 & 1049 & 1071  \\ 
No source& 5 & 26 & 31 \\ 
\hline 
Total& 108 & 1441 & 1549\\ 
\hline
\end{tabular}
\caption{\label{tab:Precisionconfusionmatrix} We use the function defined in Eqn. ~\ref{eqn:exponential_boundary} to segment the enriched sample to identify anomalous frames and show these predictions in terms of truth labels derived based on zoo consensus. Results show only a single false negative for the Anomalous class discussed further in section \ref{sec:errors} }
\end{table}

\begin{table}
\begin{tabular}{l c c c}
\\
\hline \hline
 Non-bias sample & Prediction: anomaly &  Prediction: other & Total \\ 
 \hline 
Zoo: Anomaly & 6 & 1 & 7  \\ 
Zoo: other & 59  & 1397 & 1456  \\ 
\hline 
Total& 65 & 1398 & 1463\\
\hline 
\end{tabular}
\caption{\label{tab:Recallconfusionmatrix} We use the function based boundary to segment the non-bias sample and construct a confusion matrix based on the binary classification of anomalies vs all other classes where zoo consensus was reached. We use this binary classification scheme to evaluate Recall and Informedness.}
\end{table}

\begin{figure*}
\centering
\includegraphics[width=12cm]{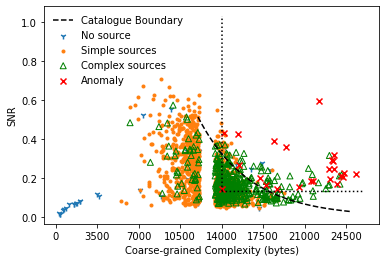}
\caption{\label{fig:Enriched_Scatter} Scatter plot showing the coarse-grained complexity and the signal-to-noise ratio (SNR), and the relationship between them, for the EMU-PS zoo sub-sample inclusive of the enrichment sample (n=1627). The figure shows the classification of frames derived through consensus from the zoo (Anomaly in the EMU Zoo) and the function based catalogue boundary for anomalies (Catalogue Boundary) with its exponential form shown in this figure and expressed in terms of a log ratio adjustment of a lower bound complexity value in equation \ref{eqn:exponential_boundary}.}
\end{figure*}

\begin{table*}
\centering
\begin{tabular}{l l c c c c c c}
\hline \hline 
 \multicolumn{8}{c}{Catalogues created} \\
\hline 
Name & Partition boundary & Sample Size & Scan  & Recall  & Informedness & Precision & Precision \\
 & & & retained & (non-bias) & (non-bias) & (non-bias) & (enriched) \\
\hline 
Most complete & Complexity boundary (Function)  & 16,157 & 5\% & 0.86 & 0.82 & 0.09 & 0.75 \\ 
Compromise &  Function \& Complexity cut $\ge$ 15,000 bytes & 3,791 & 1\% & 0.71 & 0.69 & 0.14 & 0.88 \\
Most pure & Function \& Complexity cut $\ge$ 17,000 bytes  & 1,558 & 0.5\% & 0.43 & 0.42 & 0.17 & 0.94 \\
\hline 
\end{tabular}
\caption{\label{tab:catsum} Informedness and Precision measures for different anomaly catalogues. The functional form of the complexity boundary is given by equation \ref{eqn:exponential_boundary}, and plotted in figure \ref{fig:Enriched_Scatter}. Precision is calculated using both the non-bias sample with the positive class based on anomalies only (complementary with the calculation of Recall) and also the enriched sample with the positive class based on both anomalies and complex objects (to evaluate the concentration of complex objects amongst the false positives retained)}.
\end{table*}

Figure \ref{fig:Enriched_Scatter} shows both the coarse-grained complexity and the SNR \footnote{The SNR is calculated as the reciprocal of the coefficient of variation for the channel intensity values as outlined in S19 and section \ref{sec:complexity_calc}. As frames contain predominantly channel values close to zero, these ratios are typically well below unity.}, and the relationship between them, for the EMU-PS zoo sub-sample inclusive of the enrichment sample (n=1627). The figure also shows the classification of frames derived through consensus from EMU in the Anomaly Zoo. We see that frames evaluated as containing an anomalous object by consensus are measured as having larger coarse-grained complexity and SNR values, with all anomalous objects having complexity values greater than  14,000 bytes and an SNR ratio greater than 0.14. The figure also shows that the concentration of frames with complex and anomalous zoo labels increases in the tail above the 99.5th percentile, a complexity value of approximately 17,000 bytes.

Table \ref{tab:miscresults} shows that Recall in the zoo sub-sample above a complexity threshold of 14,000 bytes is 1.00 , equivalent to 100\% retention of anomalous objects, however the false positive rate of 0.37 results in an Informedness of 0.63. The false positive rate here is measured based on the proportion of frames not containing anomalous objects by consensus that fall above the threshold. The results show that when an SNR threshold of 0.14 is incorporated, the Recall remains at 1.00, however the false positive rate reduces to 0.17 resulting in an Informedness of 0.83. The rationale for including the SNR is discussed further when considering classification errors in section \ref{sec:errors}.

In order to segment frames containing anomalous sources, and reduce the false positive rate, we fit (based on visual inspection) a function as the catalogue boundary in terms of the complexity $C(f(x))$ and SNR. This is illustrated in Fig.~\ref{fig:Enriched_Scatter} as an exponential curve where truth labels were defined using consensus votes from zoo classifications. The functional catalogue boundary can also be expressed in terms of a log ratio adjustment of a lower bound complexity value for anomalous frames
\begin{equation} 
	C(f(x)) \ge 13101\ln\left(\frac{2}{\sqrt[3]{\mathrm{SNR}}}\right),
	\label{eqn:exponential_boundary}
\end{equation}
To be considered a candidate for the anomaly catalogue, a frame must have a complexity value greater than the right side of Eqn. ~\ref{eqn:exponential_boundary}. This equation provides a complexity boundary of $C(f(x))>14000$ when the SNR of the frame is roughly 0.3. The majority of frames have a SNR less than 0.3 (as we can see from Fig.~\ref{fig:Enriched_Scatter}), and so would require a higher complexity to be considered as part of the anomaly catalogue. Using a function based catalogue boundary that incorporates both complexity and the SNR improves the Precision and Informedness of the segmented sample by reducing the retention of false positives, as illustrated in figure \ref{fig:Enriched_Scatter}. We propose that a similar partition will be effective with respect to the full EMU survey.

We use the function defined in Eqn. ~\ref{eqn:exponential_boundary} to segment the enriched sample to identify anomalous and complex frames and show these predictions in terms of zoo labels in Table \ref{tab:Precisionconfusionmatrix}. We measure the Recall, Informedness and Precision based on the binary classification of anomalies vs all other classes using the non-bias sample with results presented as a confusion matrix shown in Table \ref{tab:Recallconfusionmatrix}. The Recall and Informedness can be evaluated with respect to the horizontal totals with results showing a Recall of 0.86 and Informedness of 0.82 for positive predictions. We use estimates of Recall and Informedness based on the non-bias sub-sample to provide more accurate estimates by avoiding the allocation of too much weight to the tail through the inclusion of the Enrichment sample.

As shown in Table \ref{tab:zooclassification}, the number of anomalies is very small compared to the number of frames containing complex objects. Accordingly, even at a very low false positive rate of 0.04 for the functional boundary, the Precision remains low as the boundary permits more familiar complex objects that are not deemed anomalies. We use the combined complex and anomalous classes to verify that the precision remains high with respect to both theses classes. We use estimates of Precision based on the enriched sub-sample to provide better coverage of the right tail of the complexity distribution, as discussed in section \ref{sec:Sub-sampling}, with the aim of providing a better representation of the concentration of complex and anomalous sources above the complexity threshold defined using the function based boundary.

The Precision can be evaluated with respect to the vertical totals of the confusion matrix returning a low value of 0.09 when computed using the non-bias sample with the positive class defined using anomalies only. Based on the enriched sample, with the positive class including both anomalies and complex objects, the Precision is estimated to be 0.75 for the functional boundary. Imposing a stricter complexity cut of 17,000 bytes the Precision is estimated to be 0.94.

 Summary statistics are presented in table \ref{tab:catsum}. Results show only a single false negative for the Anomalous class discussed further in section \ref{sec:errors}. We note that the Informedness of the function based catalogue boundary is comparable to the orthogonal thresholds, made at a complexity cut of 14,000 bytes and an SNR cut at 0.14 as shown in table \ref{tab:miscresults}. A key advantage of the function based catalogue boundary however is the much smaller false positive rate with respect to anomalies, less than one quarter ($<25\%$) the size. This results in a much more efficient search space for unusual objects and can be used to produce an anomaly catalogue of a more manageable size. Imposing stricter complexity thresholds on top of the functional boundary can further reduce the false positive rate and improve Precision as shown in table \ref{tab:catsum}. These results also show a significant improvement in Precision with respect to both anomalies and complex objects can be achieved when stricter complexity thresholds are overlaid, with results showing an enriched Precision of 0.88 and 0.94 when applying complexity thresholds of 15,000 bytes and 17,000 bytes respectively. These results allow the trade-off between catalogue size, Recall and Precision to be evaluated for different boundary choices. 

\section{Discussion}
\label{sec:Conclusions}

\subsection{Anomaly Catalogue}

We construct three anomaly catalogue partitions, as shown in Table \ref{tab:catsum}, the first based only on a function incorporating complexity and SNR values to formulate the boundary, and the other two combining the function based catalogue boundary with a complexity cut at $\ge$15,000 bytes and $\ge$17,000 bytes. 

Using the function based catalogue boundary, without any further complexity cuts, produces a partitioned sample with the highest Recall of 0.86. However, it also produces a relatively large catalogue of n=16,157 frames with an estimated false positive rate of 0.04. 

To provide a more efficient search space that promotes the discovery of unusual and novel objects, we introduce further complexity cuts to reduce the catalogue size. At a complexity cut of $\ge$17,000 bytes, the catalogue size is reduced significantly from 16,157 frames to 1,558 frames with an estimated false positive rate of 0.01. We note however, that despite the reduction in the false positive rate, the Precision remains low at 0.17 with respect to the positive class based on anomalies only. Evaluating the precision based on the Enriched sample and including both anomalies and complex objects within the positive class we achieve a precision of 0.94 suggesting that the majority of objects captured in this catalogue will have some structure, and will be more interesting than simple unresolved sources or single component sources. We expect that very few of the frames belonging to this catalogue would be classified as containing only simple objects or noise. This is the smallest catalogue, and provides a fast search space, minimising the number of false positives, however the low Recall of 0.43 presents a significant risk that new and interesting objects of interest will not be captured.

At a compromise complexity cut of $\ge$15,000 bytes, the catalogue size of n=3,791 remains significantly smaller than the catalogue when excluding a complexity cut (n=16,157). The estimated Recall of 0.71, Informedness of 0.69 (based on a false positive rate of 0.02), Precision of 0.14 and enriched Precision of 0.88 (based on more familiar complex objects and anomalies in the positive class) also provide a middle ground, combining improved coverage of actual anomalies with a high concentration of interesting objects, albeit both complex and anomalous. Furthermore, the Most Pure sub-sample can be extracted from the Compromise sub-sample by filtering above a complexity of 17000 bytes. 

The Compromise catalogue is made available as supplementary material. Using this catalogue we extract 36 examples of high complexity objects, each with a complexity of 17,000 bytes or greater, and present these in Appendix \ref{appendix:examples}. Here we discuss in more detail example objects of interest, concentrating on the more unusual features of the objects, and we group these together in the currently most commonly used radio-morphological classification schemes.

\subsection{Using the Anomaly Catalogue}
\label{sec:usage}

The catalogue contains the complexity value, the SNR and the sky co-ordinates of each frame centre (RA DEC in deg). As the frames represent a square aperture of approximately $\sim$12 arcmin it is recommended that the co-ordinates of each frame centre be used to conduct a search for objects and structures of interest within a radius of 9 arcmin. It is also important to note that as frames are sampled with a stride of one quarter the frame length, many frames will be associated with the same objects of interest, with many directly over-lapping. To help make the search more efficient, frames have been grouped by clustering frame centre locations based on the distance between them. The cluster number associated with each frame is provided in the catalogue. This can help improve the efficiency of the search once an apparent host object of interest has been located within a cluster. The catalogue also contains an indicator variable to identify high complexity frames positioned adjacent to the edges of the EMU-PS region. This is to help identify frames where contributions to complexity may come from incidental structure or artefacts produced near the edges. These frames have not been assigned a complexity cluster but have been retained in the catalogue as they may still be associated with objects of interest.

\subsection{Classification errors}
\label{sec:errors}

The evaluation of classification errors provides important context for selecting a partition boundary for an anomaly catalogue. Type II errors, representing the incorrect classifications of true positives (i.e. false negatives), remained low at the boundaries evaluated. All anomalies identified by zoo participants were contained by orthogonal thresholds, made at a complexity of 14,000 bytes and an SNR of 0.14. The function based catalogue boundary produced only one false negative as shown in figure \ref{fig:Enriched_Scatter} (below the perforated exponential line).

\begin{figure}
\includegraphics[width=8cm]{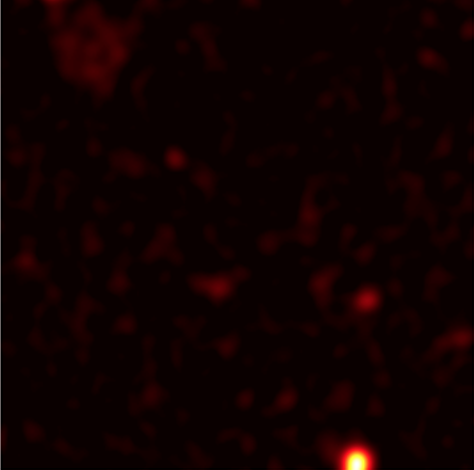}
\caption{\label{fig:FN_source} False negative errors may arise due to the removal of meaningful information by the smoothing functions due to the faint representation of complex features. This was the case with the single false negative falling outside of the function based catalogue boundary. A faint complex structure is apparent in the top left corner of the frame. While consensus converged on the classification of this frame as an anomaly, 3 out of 7 zoo participants evaluated the frame as containing only simple objects (most likely referring the bright source at the bottom of the frame towards the right side). In alternative overlapping frames, outside of the zoo sub-sample, even larger complexity values are attributed to this structure.Further investigation shows that the faint complex radio emission in the top left corner of the frame appears to be due to a face-on spiral galaxy ESO 236-G008 (z=0.03088, optical angular diameter $\sim1.5'$). Interestingly, the complexity value separates this frame from even more unusual faint radio structures such as ORC-like ringed radio emissions.}
\end{figure}

False negative errors may arise due to the removal of meaningful information by the smoothing functions, potentially due to the sparse or faint representation of complex features, discernible to the human eye or through additional measurements, but having a reduced impact on the information content of the frame itself. This was the case with the single false negative falling outside of the function based catalogue boundary. This example is illustrated in figure \ref{fig:FN_source}, where a faint complex structure is apparent in the top left corner of the frame. While consensus converged on the classification of this frame as an anomaly, 3 out of 7 zoo participants considered the frame as not containing any complex structures, and evaluated the frame as containing only simple objects (most likely referring to the bright source at the bottom of the frame towards the right side). Even though this frame falls outside of the function based catalogue boundary, a complexity value of over 14,000 bytes suggests that this frame does contain a complex structure. In alternative overlapping frames, outside of the zoo sub-sample, even larger complexity values are attributed to sections of sky containing this structure. Further investigation shows that the faint complex radio emission in the top left corner of the frame appears to be due to a face-on spiral galaxy ESO 236-G008 (z=0.03088, optical angular diameter $\sim1.5$ arcmin). Interestingly, the complexity value of this frame falls below that of, and separates it from, even more unusual faint radio structures such as ORC-like ringed radio emissions. 

 False negative errors can also be caused by the mislabelling of noise and simple objects as positives (i.e. incorrect assignment of truth labels) due to the judgement error of zoo participants. Errors can also be caused by a failure of zoo participant evaluations to converge accurately given the small number of potential volunteers with the appropriate expertise to evaluate certain objects.

\begin{figure}
\includegraphics[width=8cm]{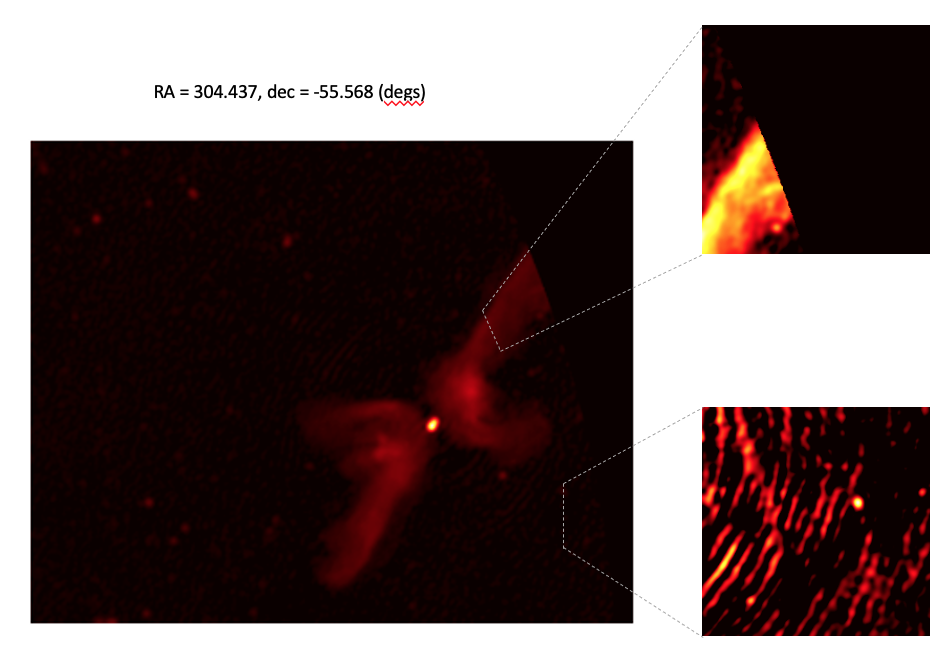}
\caption{\label{fig:xshaped} The X-shaped radio galaxy 2MASX J20180125-5539312 provides an example where frames captured only part of the larger structure. These partial frames were sometimes misclassified by participants as not being of interest.}
\end{figure}

Type I errors, representing the incorrect identification of simple sources as complex (i.e. false positives), may be due to the presence of non-random information deemed by a human observer to be incidental and not contributing to the complexity of the source itself. An example could be a telescope imaging artefact containing structure, such as a point spread function originating from a brighter source. Type I errors may also occur where multiple simple sources are captured within a frame, as the number of sources captured in each frame can vary significantly. 

Errors again may also be due to a failure of volunteer evaluations to converge on accurate truth labels, in this case mislabelling frames containing complex and unusual  objects as negatives potentially due to the loss of information in each frame due to the sampling process. This is particularly likely where only part of an interesting source was captured in the sample frame. For example, the X-shaped radio galaxy 2MASX J20180125-5539312, as shown in figure \ref{fig:xshaped}, is large enough  that the scan frames capture only part of this larger structure. These examples were misclassified by some participants as frames containing only noise or simple objects. While expert astronomers did in fact identify these frames as containing complex and anomalous structures, misclassifications by some volunteers resulted in the failure of some of these frames to reach a consensus vote that converged on a positive label. Furthermore, the human criteria for what is interesting or anomalous may be somewhat tarnished by previous exposure and not judged on its own merits (e.g. a galaxy image may have a high complexity, but be identical to images previously seen by the human judge and accordingly classified as simple). 

Alternatively, type I errors may be explained by the retention of random inputs not removed by the smoothing function. In S19 we demonstrated that in the ATLAS sample there was a large amount of noise in the simple sources at baseline. This presents a risk that in some images random inputs will take the form of incidental structure that may not be removed as smoothing increases. Where random inputs are retained after smoothing, segmentation efficiency is likely to be improved by incorporating thresholds in both the coarse-grained complexity and the SNR, as incidental structure and imaging artefacts are less likely to have the same gradient structure as astronomical sources and are therefore likely to be distributed more uniformly across the available channel values. Figure \ref{fig:Enriched_Scatter} illustrates the benefit of incorporating the SNR to reduce false positives (type I errors).

\section{Summary}
\label{sec:summary}
The coarse-grained complexity can be used as a tool for identifying unusual and complex images, useful for segmenting complex images from simple images, as demonstrated in S19. In this work we apply it to the new and larger EMU-PS data (365,000 sampled frames containing at least the $\sim$220,000 {\it Selavy} catalogue sources) after being calibrated on the much smaller ATLAS DR1 data used in S19 ($\sim$700 training sources) to identify and segment unusual sources (i.e. anomalies). This demonstrates the generalisability of the coarse-grained complexity for identifying unusual sources in increasingly larger deep radio continuum surveys such as the full EMU survey. The supporting steps taken and results discussed in this paper are summarised as follows:

\begin{itemize}

\item We scanned the mosaic image of Pilot Survey of the Evolutionary Map of the Universe, measuring the complexity of frames rather than individual sources, and examined the distribution of complexity values. An important feature of the scan method is that the frames are sampled from the EMU-PS data in a blind manner, without reference to any source catalogue.

\item We found that the high-value tail of the complexity distribution comprises many unusual, complex and extended objects. Extended sources featured heavily in the tail above the 95th percentile, with wide angle tail radio galaxies and objects of a more anomalous appearance apparent from above the 99th percentile. The ability of the coarse-grained complexity measure to segment these sources in the distribution tail shows the effectiveness of this approach at identifying scientifically interesting observations in large data sets in an efficient manner. 

\item These results demonstrate the effectiveness of this approach in recovering extended or complex structures. Examples are presented in Appendix \ref{appendix:examples}. These include peculiar FRII and FRI type sources, Tailed Radio Galaxies, Odd Radio Circles, and Giant Radio Galaxies including a previously unreported Giant Radio Galaxy with largest linear scale (LLS) of 1.94 Mpc shown in panel 5d of figure \ref{fig:complexexamples}. 

\item We used a Zooinverse project to produce crowd-sourced labels for a sub-sample of frames produced by a blind scan of the EMU-PS data to evaluate the effectiveness of the coarse-grained complexity as an anomaly partition. We identified an effective anomaly partition using the coarse-grained complexity and SNR values. These can be used to generate an anomaly catalogue and we propose that a similar partition will be effective with respect to the full EMU survey. 

\item We generate three anomaly catalogue boundaries using coarse-grained complexity and SNR values to partition the frames. The most complete catalogue boundary uses a function based on complexity and SNR values to formulate the boundary, and has n=16,157 frames, less than 5\% of the total n=365,000 frames produced from the EMU-PS scan, with an estimated Recall of 0.86 and a Informedness of 0.82. 

\item The most pure catalogue boundary has an additional complexity cut at $\ge$17,000 bytes (on top of the functional boundary), with a catalogue size of 1,558 frames and an estimated Recall of 0.43 and Informedness of 0.42. While Precision is measured at 0.17 based on anomalies only, by redefining the positive class in terms of both familiar complex objects and anomalies this partition returns an enriched Precision of 0.94.

\item The compromise catalogue boundary uses a lower complexity cut of only $\ge$15,000 bytes and has a catalogue size of n=3,791, with a Recall of 0.71 and Informedness of 0.69, providing a middle ground with respect to catalogue size and Recall. We make this catalogue available to the community as supplementary material.

\end{itemize}

The analysis has demonstrated the ability of the coarse-grained complexity to single out regions of the sky that contain complex and unusual sources based on calibration from a much smaller sample. The approach is efficient to compute and reduces expectation bias as it can be computed at worst-case linear time complexity without reference to the broader sample or existing catalogue data. This positions the approach as an efficient and powerful tool in identifying new and anomalous sources in the full EMU survey, as well as subsequent large and deep radio continuum and optical imaging surveys.




\section*{Acknowledgements}

We thank all the members of the radio astronomy community who provided their time and expertise in classifying the images as part of the Anomaly in the EMU Zoo Zooniverse.org project.
DP is supported by the project \begin{CJK}{UTF8}{mj}우주거대구조를 이용한 암흑우주 연구\end{CJK} (``Understanding Dark Universe Using Large Scale Structure of the Universe''), funded by the South Korean Ministry of Science. ELA gratefully acknowledges support from the UK Alan Turing Institute under grant reference EP/V030302/1 and from the UK Science \& Technology Facilities Council (STFC) under grant reference ST/P000649/1. H.A. benefited from grant CIIC 138/2022 of DAIP, Universidad de Guanajuato, Mexico. HT gratefully acknowledges the support from the Shuimu Tsinghua Scholar Program of Tsinghua University. JP acknowledges support from the Science and Technology Facilities Council (STFC) via grant ST/V000624/1.

The Australian SKA Pathfinder is part of the Australia Telescope National Facility (\url{https://ror.org/05qajvd42}) which is managed by CSIRO. Operation of ASKAP is funded by the Australian Government with support from the National Collaborative Research Infrastructure Strategy. ASKAP uses the resources of the Pawsey Supercomputing Centre. Establishment of ASKAP, the Murchison Radio-astronomy Observatory and the Pawsey Supercomputing Centre are initiatives of the Australian Government, with support from the Government of Western Australia and the Science and Industry Endowment Fund. We acknowledge the Wajarri Yamatji people as the traditional owners of the Observatory site.

This publication uses data generated via the Zooniverse.org platform, development of which is funded by generous support, including a Global Impact Award from Google, and by a grant from the Alfred P. Sloan Foundation.

This research has made use of the NASA/IPAC Extragalactic Database, which is funded by the National Aeronautics and Space Administration and operated by the California Institute of Technology.

This research has made use of the VizieR catalogue access tool, CDS, Strasbourg, France (DOI: 10.26093/cds/vizier). We acknowledge the usage of the HyperLeda database (http://leda.univ-lyon1.fr).


\section*{Data Availability}

The EMU Pilot Survey radio continuum stokes I image used in this analysis described in section \ref{sec:emupilotdata} was generated from data available from the CSIRO ASKAP Science Data Archive (CASDA). 

We provide the anomaly catalogue based on the compromise catalogue
boundary, as described in section \ref{sec:Conclusions}, as supplementary material. Section \ref{sec:usage} provides guidance on how to use the catalogue.



\bibliographystyle{mnras}
\bibliography{references} 




\appendix

\section{Sub-sampling exclusions}
\label{appendix:lefout}
\begin{figure}
\includegraphics[width=8cm]{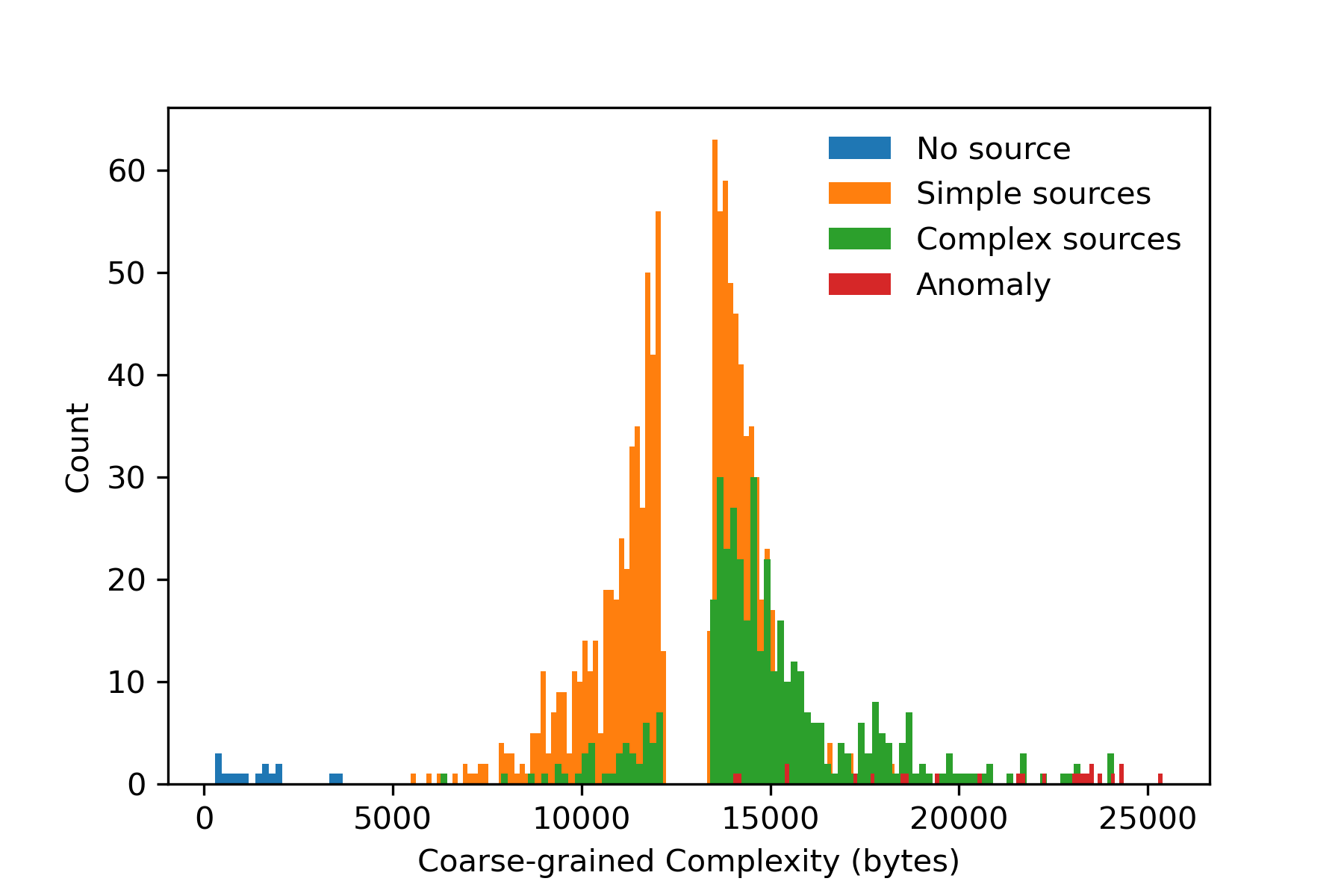}
\caption{\label{fig:4hist} Histogram partitioned by consensus votes from the zoo. Bins in the range 12,200 bytes to 13,400 bytes were omitted as described in appendix \ref{appendix:lefout}.}
\end{figure}

\begin{figure}
\includegraphics[width=8cm]{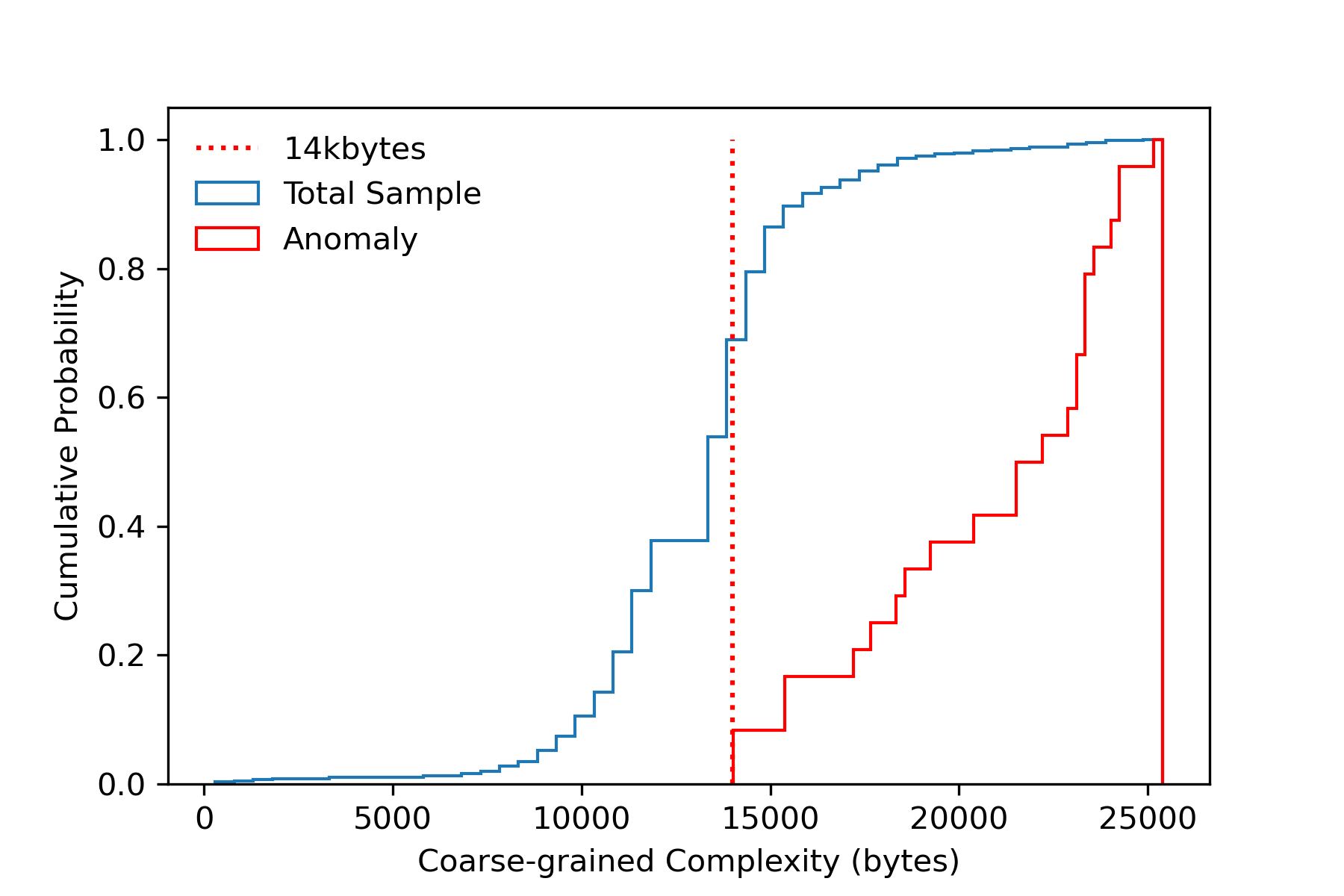}
\caption{\label{fig:Cumulative_Enriched} Cumulative probability of complexity values from both the Total sample and for anomalies identified in the zoo}
\end{figure}

Figure \ref{fig:4hist} shows the histogram of complexity values from the zoo sub-sample. Complexity values ranging from 12,200 bytes to 13,400 bytes were not included in the zoo sub-sample due to the selection procedure of inward sampling from the tails. Values within this range occur within the inner quartiles of the EMU-PS sample. The zoo results, as depicted in figure \ref{fig:Cumulative_Enriched}, show that all sources classified by zoo participants as unexpected or anomalous have a measured complexity value of approximately 14,000 bytes or greater. It is assumed that the frames not sub-sampled within the complexity range 12,200 bytes to 13,400 bytes would have minimal impact on the assessment of Recall, given the low probability of sources being classified as unexpected or anomalous below 14,000 bytes. Any false negatives (Type II errors) resulting from defining the partition boundary at this level or above are expected to be few. This is supported by the large zoo sub-sample below 14,000 bytes, over 60\% of the total zoo sub-sample, in which no sources were classified as anomalous. Furthermore the fraction of anomalous sources appears to reduce quickly at lower complexity values, as demonstrated by the steep slope at high complexity values in the cumulative probability distribution for anomalies, as shown in figure \ref{fig:Cumulative_Enriched}. Accordingly, re-sampling from within the missing range was not deemed as necessary. 

\section{Performance evaluation}
\label{appendix:evaluation}

\begin{table*}
\centering
\begin{tabular}{ c  c  c  c }
\\
\hline \hline
 & \textbf{Class +} &  \textbf{Class -} & \textbf{Total} \\ \cline{1-4} 
\textbf{Prediction +} & TP & FP (Type I error) & Predicted Positives (PP) \\ \cline{1-4}
\textbf{Prediction -} & FN (Type II error) & TN  &  Predicted Negatives (PN) \\ \cline{1-4}
\textbf{Total} & Real Positives (RP) & Real Negatives (RN)  &  \\ \cline{1-4}
\end{tabular}
\caption{\label{tab:confusionmatrix}Confusion Matrix for binary classification problem.}
\end{table*}

\subsection{Recall and Precision}

To measure Recall and Precision using the results from the Anomaly Zoo, we use the following metrics: 

\begin{equation}
    \mathrm{Recall}~=\frac{TP}{TP+FN}\,.
\end{equation}

\begin{equation}
    \mathrm{Precision}~=\frac{TP}{TP+FP}\,,
\end{equation}

Here FP are False Positives (or Type I errors, objects misidentified as being of interest), FN are False Negatives (or Type II errors, complex and unusual objects misidentified as being simple), TP are True Positives (correct positive classifications) and TN are True Negatives (correct negative classifications).

Precision  determines the number of correct positive classifications as a fraction of all positive classifications, TP/(TP+FP), while Recall determines the number of correct positive classifications as a fraction of the total number of real positives (RP=TP+FN), and so is TP/RP. The fraction of positive objects that have been missed would be $1-{\rm Recall}$, in the binary classification case.

\subsection{Informedness}

An alternative framework for measuring performance involves the use of Receiver Operating Characteristic (ROC) curves. The use of ROC curves to construct a comparative framework has been adopted in the machine learning literature \citep{Flach05}. These approaches account for chance level performance and can also be used to account for the cost weightings of negative and positive cases. ROC analysis examines the false positive rate (FP/RN) versus the true positive rate (TP/RP), which can be used to account for the trade-off between these two measures. 

The maximum positive distance of the receiver operating characteristic (ROC) curve from the 45 degree chance line is known as Youden's J statistic \citep{Youden} or as the Informedness measure \citep{powers11}. The Informedness measure is equivalent to the subtraction of the false positive rate (FPR) from the true positive rate (TPR) as follows:

	\begin{equation} \textbf{Informedness} = \frac{TP}{TP+FN} - \frac{FP}{TN+FP} = \mathrm{TPR} - \mathrm{FPR}
\end{equation}

\citet{powers11} shows that  Informedness is an unbiased estimator of above chance performance. The measure incorporates both Type I errors (False Positives) and Type II errors (False Negatives) and describes the improved performance of the measured classifier with respect to chance, costing true positives and false positives in a way analogous to how a bookmaker fairly prices the odds \citep{powers11}. For this reason the measure is also referred to as Bookmaker Informedness. The Informedness measure is defined on a (-1,1) interval and gives equal weighting to the true positive and false positive rate.

Informedness appears appropriate for evaluating the effectiveness of alternative approaches at detecting and classifying complex and unusual observations in large astronomical data. The Informedness measure relates to the following objectives of classification:

\begin{enumerate}
	\item 	\textbf{Maximise true positive rate} (i.e. minimise the type II error rate) - ensuring the search space contains as many truly interesting things as possible.
	\item 	\textbf{Minimise false positive rate} (i.e. minimise the type I error rate) - ensuring the search space does not become too large and predominantly contains truly interesting things.
\end{enumerate} 

Removing false positives reduces the search space, and the associated costs of handling larger data sets, helping to make the discovery process more efficient. Due to the likely small number of actually unusual observations (new types of objects) compared to normal observations (objects belonging to an already known or common class) in the total sample, the metric is likely to be more sensitive to small changes in the true positive count resulting from misclassification or disagreement between zoo volunteers.

In assessing the effectiveness of the approach, a key consideration is the reduction of the type II error rate, measuring the effectiveness of the approach at identifying as many of the unusual observations as possible. Minimising the type I error rate is also of importance in providing a significant reduction in the search space. Reducing the type I error rate also reduces contamination of the search space by simple sources and noise. This is reflected in the complementary measure of Precision.  For these reasons the Informedness and Precision measures were chosen as the principle criteria for evaluating prospective partition boundaries.

\section{Examples of high complexity sources}
\label{appendix:examples}

Table \ref{tab:catsum} outlines three boundaries used to construct and partition an anomaly catalogue using a complexity cuts.  

Examples of interesting non-overlapping frames from within the catalogue boundary, with a complexity of 17,000 bytes or greater, are shown in figure \ref{fig:complexexamples}. Frames with figure reference 2C, 4A, 4B, 4C and 7A were found efficiently within the catalogue search space. The remaining frames shown in figure \ref{fig:complexexamples} were sampled from within selected fractional percentile bins above 17,000 bytes and provide a representation of the high value complexity tail and the diverse morphology found within. We found the apparent host object associated with each example frame and detail these along with their main characteristics in table \ref{tab:my_label}.

\subsection{Notes on Examples of interest}

\begin{figure*}
\includegraphics[width=18cm]{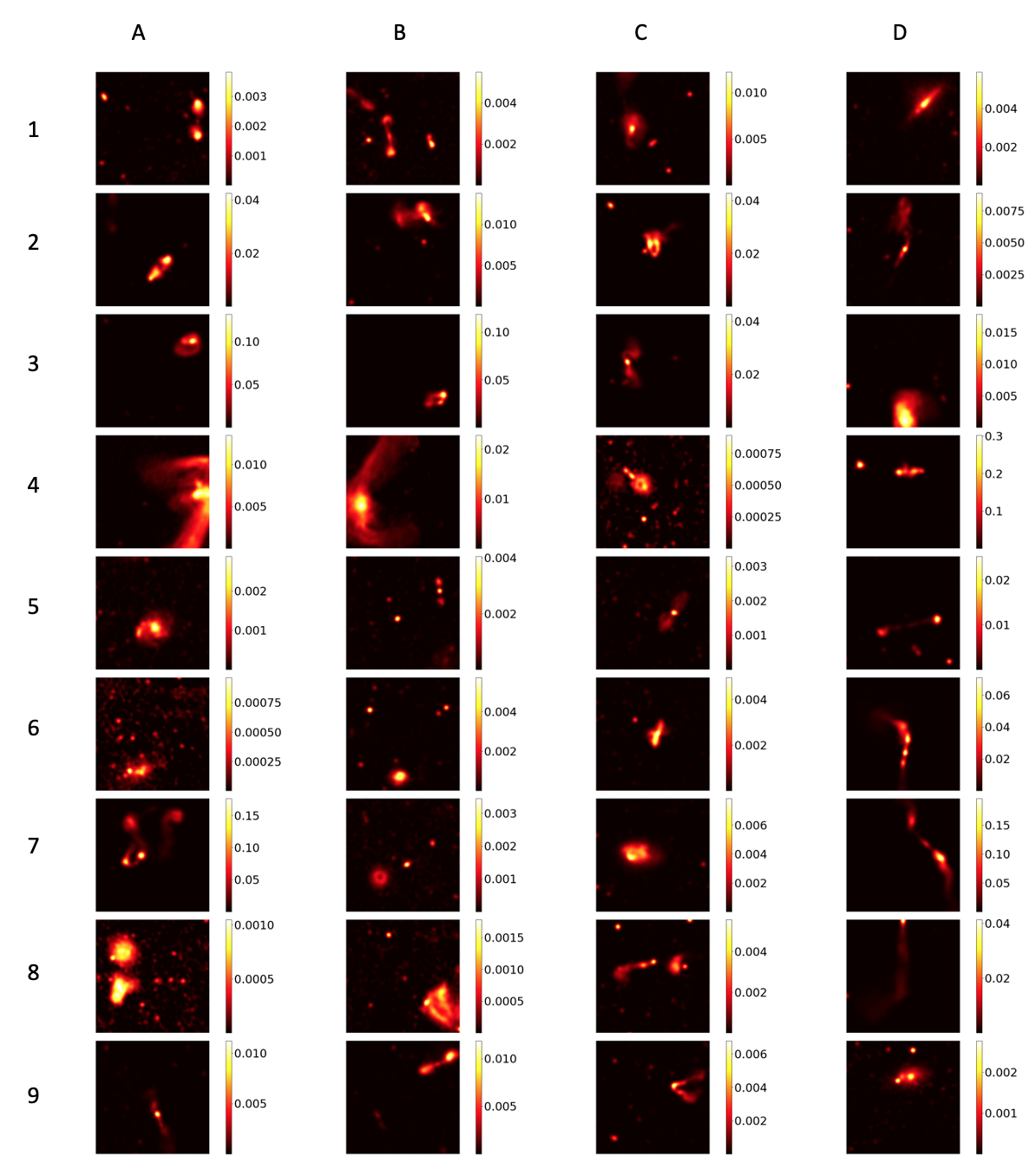}
\caption{\label{fig:complexexamples} Examples of interesting non-overlapping frames with a complexity of 17,000 bytes or greater. All frames presented are of equal size, showing a 256 $\times$ 256 pixel region (equivalent to a span of approximately $\sim 12$ arcmin). These frames were used to locate each object of interest however the associated characteristics of each object and the precise coordinates of the host were determined after the object was found. Coordinates of the optical host for each object are provided in Table~C1. The color bars show the intensity scale for each image based on the flux density in each pixel (mJy/beam).}
\end{figure*}

We looked up the regions of the 36 panels displayed in figure \ref{fig:complexexamples} 
in both the EMU PS full-resolution radio image \citep{2021PASA...38...46N} 
as well as in the deep optical images of the
\textit{Dark Energy Survey Data Release~1} (DES, \cite{2018ApJS..239...18A}).
as offered in the \texttt{Aladin} software package \citep{2000A&AS..143...33B}.
This allowed us to identify the apparent host object, and we 
retrieved their main characteristics (exact position, brightness, spectroscopic
redshifts, etc.) from the NASA/IPAC Extragalactic Database (NED, \texttt{ned.ipac.caltech.edu})
as well as the VizieR catalogue browser at CDS \citep{2000A&AS..143...23O}. 
Photometric redshifts were searched in various catalogues not readily 
available from the above data sources, like 
\cite{2014ApJS..210....9B,2016ApJS..225....5B,2021MNRAS.501.3309Z,2022RAA....22f5001Z}
and an average value was adopted if more than one was available.
The largest angular sizes (LAS) of the radio sources (including parts that
exceed the panel limits of figure \ref{fig:complexexamples}) were measured and converted to largest
linear sizes (LLS) using standard cosmological  parameters,
$H_0=70$\,km\,s$^{-1}$\,Mpc$^{-1}$, $\Omega_m$=0.3, $\Omega_{\Lambda}$=0.7.
These details along with the location of the apparent host object are provided 
in table \ref{tab:my_label}. 

In what follows we describe our findings, concentrating on the more 
unusual features of the objects, which we group together in the currently
most commonly used radio-morphological classification schemes, with 
the major such scheme being the Fanaroff-Riley class (I or II according
to \cite{1974MNRAS.167P..31F}) followed by wide-angle (WAT) or narrow-angle
tailed (NAT) radio galaxies, and less well-defined morphological classes.

\subsubsection{FR\,II type sources}

The FR\,II type radio galaxies shown in figure \ref{fig:complexexamples} tend to have more inflated
lobes than that of average objects with this classification, most likely because these features
contribute to the complexity level. The most regular and at the same time
linearly largest FR\,II is seen in panel~5D, while the
others (panels 1A, 1B, 1C, 2A, 4D, 5B, 9B) tend to have more inflated lobes of
the remnant type, the most extreme of these being the one in panel~8A,
already presented in Fig.~23 of \cite{2021PASA...38...46N}.

\subsubsection{FR\,I type sources}

FR\,I type sources are more abundant in our selection for figure \ref{fig:complexexamples}. Apart
from some sources mentioned elsewhere in this section, we emphasize the
WAT-like one in panel~2D which has two symmetric inner jets but only one 
complex lobe due North, while a South lobe or tail is not detected. 
The source's linear size, already without a South lobe, is close to 1\,Mpc. At
an adequate contrast the N jet is seen to follow a NNE direction,
then curving backward on itself in projection due SSE. The host may be
either one of two bright galaxies in a close pair, whose interaction
may be responsible for the wiggles in the jets.

\subsubsection{Tailed Radio Galaxies like WATs and NATs}

The angularly largest WAT is the one in panel~7D, whose N lobe is
seen only in the lower section of panel~3D. Another large WAT is
the one seen in both panels~6D with its central part and 8D with
its Southern tail. Salient features in the EMU PS radio image
are the sharp edges of the widening S tail and a circular shock
feature in the NNE tail. The two most peculiar WAT-type sources
are those in panels 7A and 8B. The one in 7A has been coined 
``Dancing Ghosts'' in \cite{2021PASA...38...46N} and features
two WATs in the same cluster, Abell~3785, as already discussed by
these authors (see their Fig.~21).
The one in panel~8B has its host just beyond the SW corner of
the panel, and was already shown in a more spectacular radio-optical 
overlay in Fig.\ 22 of \cite{2021PASA...38...46N}.

Another odd-shaped WAT is that in panel~3B which looks like a radio
ring but clearly has its very bright cD-like host galaxy at one
radio knot due SW. The tails are likely curved in projection along
the line of sight to cause the impression of a ring. The host is
by far the brightest galaxy towards the cluster Abell~3796, and is likely
a member of the intermediate-distance group of three (with $z\sim$0.056,
0.076 and 0.094) that seem to cause the illusion of a single rich cluster.

Panel~2C shows a WAT with its host at its southern tip. The tails bend
rather sharply and symmetrically due N before they are fairly abruptly
bent due W, perhaps due to shear in the intergalactic medium.
A very twisted, but still rather symmetric WAT is seen in panel~3C,
and another WAT rather sharply bent near its host at its Eastern tip
is shown in panel~9C.

No actual NATs are among those in figure \ref{fig:complexexamples}, 
possibly due to their complexity falling below the 99.5th percentile. 
Analysis of the complexity distribution of {\it Selavy} cutouts 
suggests NATs are more common above the 98th percentile.

\subsubsection{Circular, X-shaped, hybrid or otherwise oddly shaped Sources}

Panel~4C shows the pair of ``odd radio circles'' (ORC\,2 and ORC\,3) 
already described in \cite{norris2021a}, \cite{10.1093/mnrasl/slab041}, and
\cite{norris22}.

Apart from the giant X-shaped source PKS~2014$-$558 \citep{2020MNRAS.495.1271C}  
there is another X-shaped source with less prominent wings in panel 6C.
Panel~4D shows an E-W oriented FR\,II radio galaxy
with plumes extending NE and SW from the core and may thus be considered 
as X-shaped as well.

Hybrid morphology sources (or HyMORS, see e.g.\ \cite{2017AJ....154..253K})
are those with FR\,I morphology on one side and FR\,II on the other.
These are very rare sources and the example closest to this in figure \ref{fig:complexexamples}
may be seen in panel~8C with a 610-kpc wide radio galaxy with an 
FR\,I-type East lobe and a shorter FR\,II-type West lobe that widens with 
distance from the core and ends in a lobe with an outer boundary 
oriented perpendicular to the main source axis. The compact source 
further W of it is due to a quasar candidate.

Panel 2B shows the radio emission of the bright elliptical galaxy Fairall~106
which extends over 2~arcmin on the DES image, showing faint optical
shells.  It is the brightest of a group of four members reported by
\cite{2015A&A...578A..61D}.  Its complex and diffuse radio emission is
predominantly East of the galaxy extending over less than three times
its optical size.

The radio emission of the source in panel~3A is peaked on the brightest 
galaxy in Abell 3826B (the more distant of two groups) and has been
interpreted as due to a (currently) E-W oriented jet that has precessed
in the past in a counterclockwise direction causing a short circular
arc in the N and a longer one in the S half of the source (see middle
panel of Fig.~8 of \citet{Nikhel}.)

Panel 5C displays a double radio galaxy with highly unusual lobe shapes. These lobes appear to consist of inner spines (jets?) accompanied by parallel elongated features on each side of the spines, somewhat reminiscent of a trident, which is suggestive of a collimated backflow of jet/lobe material after reaching the outer ends of the source, and which show no prominent hotspots. While the radio core appears to be extended due W, almost
perpendicular to the general source axis, this is likely due to a
point source $\sim11''$\,WSW, on the unrelated galaxy DESI J325.8764-51.0961
which appears to be located at a similar redshift.

\subsubsection{Giant Radio Galaxies, GRG}

As can be seen from table \ref{tab:my_label}, there are five sources that exceed an LLS
of 1\,Mpc. The largest, and previously unreported, one of 1.94\,Mpc is
seen in panel 5D of figure \ref{fig:complexexamples}, is hosted by an r=20.2\,mag
galaxy. EMU PS shows continuous emission from one end of the source to the
other, suggesting the presence of jets all the way from the core to the
lobes which is unusual for such large radio galaxies. The next largest GRG
of 1.57\,Mpc (panels 4A and 4B) is the largest X-shaped source known,
PKS~2014$-$558, and was studied in detail by \cite{2020MNRAS.495.1271C}.
The source in panel 9A reminds of a restarted (or double-double) radio
galaxy with its inner radio knots on opposite sides of the host, but
the outer FR\,I-type lobes rather suggest it is a WAT oriented with
its plane containing our line of sight. However, there is no evidence
for a cluster in the DES image, and its size of 1.37\,Mpc is unusual
for WAT sources. Panel 5C shows a core-dominated double source with
trident-shaped lobes, already described above and extremely unusual for
its large LLS of 1.30\.Mpc. The fifth-largest source of 1.15\,Mpc is
seen in panel 1C, although its northern lobe reaches beyond the panel
size. It has very diffuse, remnant-type inflated lobes, again unusual for
such large sources.  Apart from these sources there are seven more
(panels 1A, 2D, 3D=7D, 4C, 5B, 6D, 8D) in the range LLS=0.7--1\,Mpc,
considered by most current authors as GRGs as well.

\subsubsection{Nearby spiral Galaxies}

Spiral galaxies, with extremely few exceptions, tend to show radio
emission that more or less extends over part or all of their optical 
extent. Edge-on spirals, when observed with sufficient sensitivity, 
may show radio emission extending away from their galactic planes.
The latter is what we see for the only perfect edge-on spiral in Fig.~C1, NGC~7090,
in panel~1D, with an inclination angle of 0$^{\circ}$ according to HyperLEDA
\citep{2014A&A...570A..13M}, where the optical galaxy is oriented SE-NW 
and the radio emission on its NE side appears more extended than on 
its SW side. Radio emission is detected over about half the projected
length of the optical disk in the DES image (see also Fig.~26 of 
\cite{2021PASA...38...46N}).
 
Panel~5A shows the radio emission of the barred and almost face-on spiral NGC~7125,
which extends over its entire optical extent, but is more patchy 
and irregular than the optical emission (see Fig.~27 of \cite{2021PASA...38...46N}).

The SE corner of panel~6A shows the amorphous radio emission of the 
dwarf irregular galaxy IC~5152, a member of the Local Group.

Panel~6B shows the barred spiral NGC~6984, which is too small
for EMU PS to resolve details of the spiral structure.

The SE quadrant of panel~7B features the ring-like radio emission of 
NGC~6935, a very high surface brightness barred spiral surrounded by 
faint and tightly bound spiral arms in the DES image, where radio emission
is still faintly seen in EMU PS.

The radio emission in panel~7C traces the spiral structure of the Sbc
type galaxy NGC~7205 of intermediate inclination.

Finally, the upper half of panel~9D shows the barred spiral NGC~7059 
with its inner region bright in both radio and optical. The compact
radio source in the Eastern outskirts of the galaxy coincides with the
X-ray source 2SXPS~J212729.2$-$600102 and is likely the counterpart 
of the $\gamma$-ray source 4FGL~J2127.6$-$5959 \citep{2021AJ....161..154K}
in the background of  NGC~7059.

\begin{table*}
    \centering
    \begin{tabular}{ccccccccl} \hline
Fig ref & Complexity (bytes) & Percentile & RAJhost & DECJhost &  LAS/' & z~~~~ztype & LLS/Mpc & ~~~~~Host Name \\ 
  (1)   &   (2)   &   (3)   &   (4)   &   (5)   &   (6)   &   (7)   &   (8)   &  ~~~~~~~~~~ (9)   \\ \hline
1A & 17348 & 99.5 & 306.8107 & $-$55.3130 &  2.88 & 0.36~~~~~~ p & 0.87  & DES J202714.57$-$551846.9   \\
1B & 18932 & 99.9 & 319.1425 & $-$63.0186 &  2.82 & 0.25~~~~~~ p & 0.66  & 2MASX J21163413$-$6301068 \\
1C & 19749 & 99.9 & 326.0088 & $-$48.3159 &  7.95 & 0.135~~~~ p & 1.15  & 2MASX J21440210$-$4818581    \\
1D & 19670 & 99.9 & 324.1203 & $-$54.5573 &  4.0~ & 0.002825 s & 0.014 & NGC 7090, edge-on spiral galaxy \\
2A & 18965 & 99.9 & 306.6852 & $-$55.3743 &  2.15 & 0.26~~~~~~ p & 0.51  & DES J202644.45$-$552227.3  \\
2B & 19467 & 99.9 & 323.7978 & $-$62.0786 &  3.6~ & 0.05629~ s & 0.24  & 2MASX J21351149$-$6204432   \\
2C & 19316 & 99.9 & 322.3293 & $-$50.8844 &  3.2~ & 0.07999~ s & 0.29  & 2MASX J21291901$-$5053040  \\
2D & 19476 & 99.9 & 325.7994 & $-$61.4718 &  5.2~ & 0.1825~~ s & 0.96  & 2MASX J21431182$-$6128184  \\
3A & 19023 & 99.9 & 330.1005 & $-$56.1782 &  1.6~ & 0.07578~ s & 0.14  & 2MASX J22002408$-$5610413  \\
3B & 17367 & 99.5 & 324.8748 & $-$51.3955 &  1.7~ & 0.07573~ s & 0.15  & 2MASX J21392989$-$5123440  \\
3C & 19987 & 99.9 & 309.1440 & $-$57.6233 &  3.56 & 0.03586~ s & 0.15  & ESO 143$-$G035, Fairall 74 \\
3D & 19152 & 99.9 & 327.8747 & $-$55.3369 & 15.2~ & 0.03878~ s & 0.70  & 2MASX J21512991$-$5520124   \\
4A & 19839 & 99.9 & 304.5055 & $-$55.6586 & 22.4~ & 0.0606~~ s & 1.57  & 2MASX J20180125$-$5539312   \\
4B & 19745 & 99.9 & 304.5055 & $-$55.6586 & 22.4~ & 0.0606~~ s & 1.57  & 2MASX J20180125$-$5539312   \\
4C & 17570 & 99.5 & 314.7033 & $-$57.6033 &  2.68 & 0.31~~~~~~ p & 0.73  & DES J205848.79$-$573612.0   \\
4D & 19629 & 99.9 & 335.2748 & $-$50.3070 &  2.2~ & 0.32~~~~~~ p & 0.62  & DES J222105.94$-$501825.2   \\
5A & 18870 & 99.9 & 327.3166 & $-$60.7131 &  3.1~ & 0.010501 s & 0.041 & NGC 7125, face-on spiral galaxy \\
5B & 17992 & 99.8 & 308.5346 & $-$52.8950 &  1.68 & 0.70~~~~~~ p & 0.72  & DES J203408.30$-$525341.8 \\
5C & 17212 & 99.5 & 325.8806 & $-$51.0948 &  3.3~ & 0.58~~~~~~ p & 1.30  & DES J214331.34$-$510541.1  \\
5D & 17867 & 99.8 & 334.1668 & $-$62.8782 &  4.86 & 0.596~~~~ p & 1.94  & DES J221640.02$-$625241.6  \\
6A & 17419 & 99.6 & 330.6730 & $-$51.2964 &  3.7~ & 0.000407 s & 0.002 & IC 5152; ESO 237$-$G027  \\
6B & 17655 & 99.7 & 314.4749 & $-$51.8708 &  1.6~ & 0.015577 s & 0.031 & NGC 6984, spiral galaxy \\
6C & 17411 & 99.6 & 324.5352 & $-$59.6218 &  1.93 & 0.21~~~~~~ p & 0.40  & DES J213808.45$-$593718.4  \\
6D & 20230 & 99.9 & 311.4677 & $-$51.1074 & 13.9~ & 0.04849~ s & 0.79  & 2MASX J20455226$-$5106267  \\
7A & 20179 & 99.9 & 323.5737 & $-$53.6363 &  4.4~ & 0.07945~ s & 0.39  & 2MASX J21341775$-$5338101  \\
7B & 17642 & 99.7 & 309.5843 & $-$52.1104 &  1.5~ & 0.015154 s & 0.027 & NGC 6935, ringed face-on spiral galaxy \\
7C & 18983 & 99.9 & 332.1429 & $-$57.4426 &  3.8~ & 0.005623 s & 0.027 & NGC 7205, spiral galaxy \\
7D & 20670 & 99.9 & 327.8747 & $-$55.3369 & 15.2~ & 0.03878~ s & 0.70  & 2MASX J21512991$-$5520124  \\
8A & 18944 & 99.9 & 312.1568 & $-$49.1876 &  4.76 & 0.11~~~~~~ p & 0.57  & 2MASX J20483764$-$4911157  \\
8B & 18906 & 99.9 & 310.3001 & $-$52.9605 &  5.8~ & 0.04801~ s & 0.33  & 2MASX J20411202$-$5257379  \\
8C & 18185 & 99.8 & 328.8026 & $-$59.1505 &  5.0~ & 0.115~~~~ p & 0.61  & 2MASX J21551267$-$5909011  \\
8D & 20215 & 99.9 & 311.4677 & $-$51.1074 & 13.9~ & 0.04849~ s & 0.79  & 2MASX J20455226$-$5106267  \\
9A & 17886 & 99.8 & 307.0755 & $-$49.4023 &  5.14 & 0.31~~~~~~ p & 1.37  & DES J202818.12$-$492408.4  \\
9B & 19108 & 99.9 & 334.8894 & $-$60.0305 &  3.3~ & 0.19~~~~~~ p & 0.63  & DES J221933.46$-$600149.9  \\
9C & 18270 & 99.8 & 322.1716 & $-$60.3659 &  2.61 & 0.10052~ s & 0.29  & 2MASX J21284113$-$6021568  \\
9D & 17913 & 99.8 & 321.8395 & $-$60.0146 &  3.5~ & 0.005784 s & 0.036 & NGC 7059, spiral galaxy \\ \hline
    \end{tabular}
    \caption{ Optical identifications of sources in all frames of Figure~C1,
   with a complexity of at least 17,000 bytes. Column 1 gives the frame
   position (row number and column letter) in  \ref{fig:complexexamples}, Column 1 provides the complexity in bytes of the sample frame used to find the object of interest, Column 3 provides the complexity percentile or the estimated probability of a frame having a complexity below this value, Columns~4 and 5 give the RA and DEC (J2000) of the
  optical host object, column~6 its largest angular radio size, and column~7
  the redshift, followed by its type (s for spectroscopic, p for photometric).
  Column~8 lists the largest linear size, and column~9 the name of the host
with an explicit mention in case of spiral galaxies.
    \emph{Comments:
    1B: Further source at frame centre, extended $\sim1'$ N-S, has no obvious optical identification; ~
    2B: Fairall 106, optical shells; ~
    3D: Only N half of N lobe is seen in frame 3D, rest of source is in frame 7D; ~
    4D: In W half of frame; central part shows 2MASX J22212664$-$5016453; ~
    5C: Trident-shaped lobes; ~
    6A: Dwarf irregular G in Local Group; ~
    6D: S part of source in panel 8D; ~
    7A: Also shows 2MASX J21340666-5334186 in Abell~3785; ~
    7D: Inner part of large WAT, N lobe in frame 3D; ~
    8D: N half of source in panel 6D.}
    }
    \label{tab:my_label}
\end{table*}

\bsp	
\label{lastpage}
\end{document}